\documentclass[12pt,preprint]{aastex}

\usepackage{verbatim}

\shorttitle{Alfv\'en Mode Scaling in Two Fluid MHD Simulations }

\shortauthors{BURKHART ET AL.}
\begin{document}

\title{Alfv\'enic Turbulence Beyond the Ambipolar Diffusion Scale}

\author{Blakesley Burkhart\altaffilmark{1,2}, A. Lazarian\altaffilmark{2}, D. Balsara\altaffilmark{3},  C. Meyer\altaffilmark{3}, J. Cho\altaffilmark{4}}
\affil{$^1$ {Harvard-Smithsonian Center for Astrophysics, 60 Garden St., Cambridge, MA 0213}}
\affil{$^2$ {Astronomy Department, University of Wisconsin, Madison, 475 N.  Charter St., WI 53711, USA}}
\affil{$^3$ {Department of Physics, University of Notre Dame, Notre Dame, IN 46556, USA}}
\affil{$^4$ {Department of Astronomy and Space Science, Chungnam National University, Daejeon, Korea }}

\begin{abstract}
We investigate the nature of the Alfv\'enic turbulence cascade in two fluid MHD simulations in order to determine if 
turbulence is damped once the ion and neutral species become decoupled at a critical scale called the ambipolar
diffusion scale (L$_{AD}$).
Using mode decomposition to separate the three classical MHD modes, we study the second order structure functions of the Alfv\'en mode velocity field of both
neutrals and ions in the reference frame of the local magnetic field. 
On scales greater than L$_{AD}$ we confirm that two fluid turbulence strongly resembles single fluid MHD turbulence. 
Our simulations show that the behavior of two fluid turbulence becomes more complex  on scales less than L$_{AD}$.
We find that Alfv\' enic turbulence can exist past L$_{AD}$ when the turbulence is globally super-Alfv\'enic, with the ions and neutrals forming separate cascades once decoupling has taken place.  When turbulence is globally sub-Alfv\'enic and hence strongly anisotropic with a large separation between the parallel and perpendicular decoupling scales,  turbulence is damped at L$_{AD}$.   We also
find that the power spectrum of the kinetic energy in the damped regime is consistent with a $k^{-4}$ scaling (in agreement with the predictions of Lazarian, Vishniac \& Cho 2004).

\end{abstract}
 
 \section{Introduction}
\label{intro}

Turbulence and magnetic fields are critical components of the interstellar medium (ISM) of galaxies from kiloparsec to sub-astronomical unit scales.
Magnetohydrodynamic (MHD) turbulence is a key element 
in the study of star formation and molecular cloud structure, magnetic reconnection, heat transport 
and cosmic ray propagation  (see  Elmegreen \& Scalo 2004; Ballesteros-Paredes et al. 2007; Mckee \& Ostriker 2007; Tilley \& Balsara 2006; Tilley, Balsara, \& Howk 2006;  Balsara et al. 2008; Brandenburg \& Lazarian 2013).
Despite the importance of turbulence for ISM studies, many mysteries
remain, including the nature of turbulence driving and damping scales.

The smallest scales of the turbulence cascade, including the damping scale, may play a pivotal role in the dynamics of giant molecular clouds (GMCs)
and star formation.
GMCs are partially ionized, with neutrals coupled to the magnetic
field through collisions with ions. The drift of the neutrals towards the central gravitational potential through the ionized particles tied 
to the magnetic field, known as ambipolar diffusion, is an often invoked source of dissipation of the MHD cascade 
(Zweibel \& Josafatsoon 1983;  Ciolek \& Basu 2000; Tassis \& Maouschovias 2004).
The ambipolar diffusion scale ($L_{AD}$), or the scale at which neutrals and ions decouple, 
has been thought to set the dissipation scale of turbulence in molecular clouds
and to set a fundamental characteristic scale for gravitational collapse and star formation 
(Balsara 1996; Houde et al. 2000; Klessen, Heitsch \& Mac Low 2000; Li et al. 2006; Li \& Houde 2008; Li et al. 2008; Hezareh et al. 2010; Tilley \& Balsara 2010; Meyer et al. 2014). 

The ambipolar diffusion scale can be estimated as the scale at which the Reynolds number, with diffusivity given by the 
ambipolar diffusivity, is equal to unity (Brandenburg \& Zweibel 1994, 1995; Balsara 1996; Oishi \& Mac Low 2006).  The ambipolar diffusivity  is given by
\begin{equation}
\nu_{AD}=\frac{B^2}{4\pi\rho_i\rho_n\alpha}
\end{equation}
where $\rho_i$ and $\rho_n$ are the density of the ions and neutrals, B is the magnetic field strength, and 
$\alpha$ is the frictional coupling coefficient\footnote{$\alpha$ is often denoted as $\gamma$ in the literature. However, in this work we use the notation $\alpha$ as it was used in Meyer et al. 2014 for consistency with this previous work. }   between the ions and neutrals.

The Reynolds number for ion-neutral drift is defined as:
\begin{equation}
R_{AD}=\frac{LV}{\nu_{AD}}
\end{equation}
where V is a characteristic velocity (e.g. for trans-Alfv\'enic turbulence it is the Alfv\'en speed, $V_A=\frac{B}{\sqrt{4\pi\rho_n}}$) and L=L$_{AD}$ when R$_{AD}$=1. This gives the form of the  ambipolar diffusion scale as often found in the literature:
\begin{equation}
L_{AD}=V_{A}/\alpha\rho_i
\end{equation}
More details about the derivation and significance of $L_{AD}$ can be found in a number of works including  Langer (1978); Zweibel \& Josafattsson (1982), Balsara (1996), Klessen Heitsch \& Mac Low (2000), Oishi \& MacLow (2006), Li \& Houde (2008),  Hezareh et al. (2010) and  Meyer et al. (2014)

The application of ambipolar diffusion extends beyond direct studies of star formation to include
general studies of magnetic fields.  For example, Houde et al. (2000); Li \& Houde (2008),  and Hezareh et al. (2010)
have proposed that the magnetic field in the plane of the sky may be obtained
from observations via calculation of the ambipolar diffusion length scale. In light of these diverse interpretations of the meaning of $L_{AD}$ it is important to understand to what extent $L_{AD}$ is relevant to turbulence damping in partially ionized gasses. 

More generally,
ambipolar diffusion has also been proposed to damp particular families of linear MHD waves (see Balsara 1996 and ref. therein).  On scales larger than L$_{AD}$ it was predicted that two fluid turbulence acts like single fluid MHD turbulence.  In particular,
Balsara (1996) on the basis of 1D dispersion analysis, 
showed that two separate mode damping situations can occur at scales at or smaller than $L_{AD}$, which are based on the value of plasma beta (i.e.
the ratio of the gas to magnetic pressure). When the Alfven speed is greater than the sound speed, the fast and Alfven wave families are damped at or below $L_{AD}$. When the Alfven speed is smaller than the sound speed, the slow and Alfven wave families are damped.
For either high or low plasma beta, Balsara (1996) predicts that the Alfv\'enic waves should damp at $L_{AD}$ and thus the MHD cascade
should damp past the ambipolar diffusion scale. To put Balsara (1996) in context, large scale 3D MHD simulations were not possible at the time.  Tilley \& Balsara (2011)
have extended the study of Balsara (1996) to include flows with radiative effects. They find that the general analysis of the MHD wave modes that are damped remained
unchanged.

Does MHD turbulence, specifically the Alfv\'en modes,  damp at the decoupling scale $L_{AD}$? MHD turbulence is known to be different from a collection of
linear Alfv\'enic waves. Cascading rates and the anisotropy of turbulence should be
accounted for carefully before we can make a definitive conclusion about
turbulent damping in the partially ionized media.
The purpose of this paper is to address the question above numerically,
as well as to numerically test the validity the Alf\'ven mode scaling relations
of the Goldreich \& Sridhar (1995, henceforth GS95) theory for ion-neutral turbulence.
The GS95 theory was first extended to a partially ionized compressible medium 
in subsequent works by Lithwick \& Goldreich (2001, henceforth LG01) and Lazarian et al. (2004, henceforth LVC04). 
In this paper we will apply the Cho \& Lazarian (2002, 2003; henceforth CL02, CL03, respectively) MHD mode decomposition technique\footnote{A different decomposition technique based on wavelet analysis is used in Kowal \& Lazarian 2010, but
the results of the two different procedures of decomposition are similar.}
to two-fluid MHD simulations, first presented in Meyer et al. (2014), in order to investigate the behavior of the Alfv\'en modes in the ions and neutrals.
Our paper is organized as follows: in Section~\ref{sec:gs95} we review the basic scaling relations predicted by the GS95 model, in Section
~\ref{sims} we describe the numerical simulations and relevant scales, in Section~\ref{nomode} we present the structure function scalings
for the the full data cubes and the Alfv\'enic modes of the ions and neutrals for our simulations and finally in Section~\ref{discussion} we discuss our results
followed by our conclusions in Section~\ref{sec:con}.

\section{The GS95 scalings}
\label{sec:gs95}

MHD turbulence is a subject with an extended history (see book by Biskamp 2003). However, it
has been given a boost more recently with the advent of 3D MHD simulations
which have allowed for testing  theoretical predictions (see recent reviews by
Brandenburg \& Lazarian 2013; Beresnyak \& Lazarian 2014). 
The modern theory of MHD turbulence is based on the Goldreich-Sridhar (1995, GS95)
idea which was extended and tested in subsequent publications (Lazarian \& Vishniac 1999, Cho \& Vishniac 2000, Maron \& Goldreich 2001, Cho \& Lazarian
2002, 2003, henceforth CL02, CL03, respectively).  The applicability of GS95
theory to the partially ionized gas was discussed in Lithwick \& Goldreich (2001) and Lazarian et al. (2004). 
 
It has been shown numerically, that in the presence of dynamically important magnetic fields eddies become elongated along the magnetic field lines. GS95 approach to Alfv\'enic modes can be easily understood:
For the eddies perpendicular to the magnetic field, the original Kolmogorov energy scaling applies, i.e.
 $V_l\sim \l_{\bot}^{1/3}$, where $l_{\bot}$ denotes scales measured perpendicular to the local
magnetic field. 
Mixing motions induce Alfv\'enic perturbations that determine the parallel size of the magnetized eddy.  
This is the concept of {\it critical balance}  i.e. the equality of the eddy turnover time ($l_{\bot}/V_l$) and the period of 
the corresponding Alfv\'en wave $\sim l_{\|}/V_A$, where $l_{\|}$ is the parallel eddy scale and $V_A$ is the Alfv\'en velocity. 
Making use of $V_l\sim \l_{\bot}^{1/3}$, one finds the scaling relation for the parallel and 
perpendicular eddies as: $l_{\|}\sim l_{\bot}^{2/3}$. 
This represents the scale dependent tendency of eddies to become more elongated along the magnetic field lines
as the energy cascades proceeds to smaller scales and corresponds to the scaling of the slow and Alfv\'en wave anisotropy.
The power spectrum for the slow and Alfv\'en waves scales as $E\sim k_{\perp}^{-5/3}$ (see review by Brandenburg \& Lazarian 2013).
Numerical studies of scaling of compressible MHD turbulence based on the decomposition into Alfv\'en, Slow and
Fast modes were first performed in CL02 and CL03 and these results were later reconfirmed in Kowal \& Lazarian (2010) with a wavelet approach.

GS95 theory assumes the isotropic injection of energy at scale $L$ and the injection velocity equal to the Alfv\'en velocity in
the fluid $V_A$, i.e. the Alfv\'en Mach number $M_A\equiv (V_L/V_A)=1$.
The GS95 model was later generalized
for both sub-Alfv\'enic, i.e. $M_A<1$, and super-Alfv\'enic, i.e. $M_A>1$, cases (see Lazarian \& Vishniac 1999 and Lazarian 2006)
and thus the simulations used in this paper (which are sub- and super-Alfv\/enic) should be understood in this context.
In the next two paragraphs we provide a brief synopsis of the differences of single fluid turbulence with 
$M_A > 1$  and $M_A < 1$.

For $M_A > 1$ magnetic fields are not dynamically important at the largest scales and  hence 
turbulence from the driving scale ($L$) to a transition scale $L_A$ follows an isotropic cascade.  
At scales smaller than $L_A$, critical balance occurs and scale dependent anisotropy proceeds down to the dissipation range.
Scale  $L_A$ is given by:
\begin{equation}
L_A=L(V_A/V_L)^3=LM_A^{-3}
\label{alf}
\end{equation}

and the relationship between parallel and perpendicular scales that occurs at scales smaller than $L_A$ is:

\begin{equation}
l_{\|}\sim L (l_{\bot}/L)^{2/3} M_A^{-1},~~~ M_A>1,
\label{supA}
\end{equation}
 where
$\|$ and $\bot$ are relative to the direction of the local magnetic field.

Similarly, for  $M_A<1$, turbulence does not obey the GS95 scaling starting at scale $L$, but
from a smaller scale $L_{trans}$, as it transitions from weak to strong turbulence: 
\begin{equation}
L_{trans}\sim L(V_L/V_A)^2\equiv LM_A^2
\end{equation}
\label{trans}
In the range $[L, L_{trans}]$ the turbulence is  ``weak,'' meaning that it is dominated by interacting MHD waves, rather than eddies.  

For scales less than $L_{trans}$
the turbulence is eddy-like (i.e. strong) and it follows a GS95-type scalings:
\begin{equation}
 l_{\|}\sim L (l_{\bot}/L)^{2/3} M_A^{-4/3},~~~ M_A<1.
\label{subA}
\end{equation}

CL02, CL03 and  Kowal \& Lazarian (2010) focused their analysis and comparisons with theoretical predictions of single fluid MHD turbulence simulations. 
Ions and neutrals are generally expected to behave as a single MHD fluid at length scales from $L_{trans}$ or $L_A$  to scale  $L_{AD}=V_{A}/\alpha\rho_i$.
At scales smaller than $L_{AD}$, the ion and neutral energetics
separate and MHD turbulence may dissipate (Zweibel \& Josafatsoon 1983; Balara 1996; Klessen 2000). 

All these above mentioned studies assumed that the diffusivities arising from viscosity and resistivity are the same, i.e. that the Prandtl number of turbulence in unity\footnote{The Prandtl number mentioned here is defined as Pr$=\nu/\eta$, where $\eta$ is magnetic resistivity and $\nu$ is fluid viscosity. Here we make the notational extension that Pr$_{AD}=\nu_{AD}/\eta$. Since the fluid diffusivity ($\nu$) is much smaller than $\nu_{AD}$ we expect Pr$>1$.}.
However, viscosity induced by neutrals can result in a very different regime
of turbulence. The high Prandtl number MHD turbulence was described theoretically in Lazarian et al. (2004, henceforth LVC04) with numerical simulations published in 
Cho, Lazarian \& Vishniac (2002, 2003). In this regime, magnetic and kinetic energy spectra are different with theoretical predictions for the kinetic energy
$E_K\sim k^{-4}$ and magnetic energy $E_M\sim k^{-1}$ (LVC04).  These predictions agree with single fluid MHD numerical simulations and we therefore use these
studies as a touchstone for our analysis.  At the same time, it would be an oversimplification to assume that turbulence in a partially ionized gas is equivalent to high Prandtl number turbulence.
It can have some features of it, but ion-neutral damping of turbulent motions as well as the decoupling of ion and neutrals change the picture in a significant way (Lithwick \& Goldreich 2002, LVC04, Xu et al. 2014, Lazarian \& Yan 2014). Therefore
in what follows we investigate numerically how MHD turbulence evolves in the partially ionized gas.

 \section{Numerical Scheme and Mode Decomposition}
\label{sims}

We use the simulations first presented in  Meyer et al. 2014 and reference to that work for the details of the numerical setup
and provide here only the essential points for this work.

The MHD capabilities of the RIEMANN code (Balsara 1998a,b; Balsara \& Spicer 1999a,b; Balsara 2004, 2010, 2012)
have recently been upgraded to treat two-fluid MHD. In two-fluid MHD, the ambipolar effects are modeled with a 
neutral fluid obeying the isothermal Euler equations and an ionized fluid which obeys the isothermal MHD equations 
(Tilley \& Balsara 2008; Tilley, Balsara \& Meyer 2011). The two-fluid version of RIEMANN has been applied to 
astrophysical turbulence in our prior papers (Tilley \& Balsara 2010, Meyer et al. 2014). A variety of higher 
order reconstruction methods are available in our code and we used r=3 WENO reconstruction (Jiang \& Shu 1996, 
Balsara \& Shu 2000) because it represents a good compromise between accuracy and speed. The density of the 
neutrals was scaled to unity and the mean molecular weights of the neutrals and ions were given by $\mu_N=2.3$ amu 
and $\mu_i=29$  amu (corresponding to HCO+) respectively. Consequently, for a particular ionization fraction  $\chi$  , the ion density is 
given by $\rho_i=\chi\rho_N\mu_i/\mu_N$ . The magnetic field was initialized along the x-direction.  
The simulations were started with uniform density in the ions and neutrals, and they were forced via 
random Gaussian fluctuations (peaked at $k=2$ and spanning $1 \le k \le 4$) into a turbulent state so 
that the desired Mach numbers were represented in the velocity field. Once a steady state was reached, 
the simulations were continued for a few further turn over times so that statistics of the two-fluid turbulence could be gathered.

\begin{table*}
\begin{center}
\caption{Description of the simulation parameters
\label{tab:models}}
\begin{tabular}{ccccccccc}
\hline\hline
Run &  $M_{s}$   & $<M_{A}>$& $\chi$  & L$_{AD\bot}$ & L$_{AD\|}$& L$_{A}$& l$_{trans}$ & L$_{Drive}$ \\
\tableline
A1 & 3.0 &1.8&  $10^{-4}$& 58-9    &40 & 77& N/A  & 512-128 \\
A3  & 2.5 &0.7& $10^{-4}$&33-16.4 &80& N/A&  256  &512-128 \\
A6 & 2.5  &0.4& $2x10^{-4}$&11-5.3 &80& N/A&  86  &512-128 \\
\hline\hline
\end{tabular}
\end{center}
\end{table*}

In this paper, we investigate three forced turbulence simulations at resolution $512^3$.  
We set the Alfv\'en speed for the combined fluid to be 1, 3 and 6 times the sound speed. We refer to these runs as A1, A3 and A6 respectively.   
Three obvious scales of interest exist in these simulations: the turbulence driving scale (L), the ambipolar diffusion scale (L$_{AD}$) 
and the numerical dissipation scale, which begins around 10-20 grid units.
We note that in the regime of strong MHD turbulence, one must think of the ambipolar diffusion scale in both the direction parallel and 
perpendicular to the mean magnetic field, i.e.
L$_{AD\|}$ and L$_{AD\bot}$, respectively. L$_{AD\|}$ as set in the code is given by L$_{AD\|}=v_A/\alpha\rho_i$.
L$_{AD\bot}$ is given by either Equation ~\ref{supA} or ~\ref{subA} depending on if the turbulence is super-Alfv\'enic or sub-Alfv\'enic, 
respectively. 
We list values of L$_{AD\bot}$  in Table 1 for the range of driving scales, i.e. from L=512-128.  Table 1 also provides the values 
for L,  L$_{AD\|}$, 
as well as the  transition  between weak and strong turbulence (l$_{trans}$) for models A3 and A6 as given by equation ~\ref{trans} and, for 
the super-Alfv\'enic model A1, the transition scale between hydrodynamic and MHD turbulence (L$_A$) given by equation ~\ref{alf}.  
In columns two, three and four of Table 1 we list
the sonic Mach number and the volume averaged neutral-ion Alfv\'en Mach number.   

\begin{figure*}[tbhp]
\centering

\includegraphics[scale=.5]{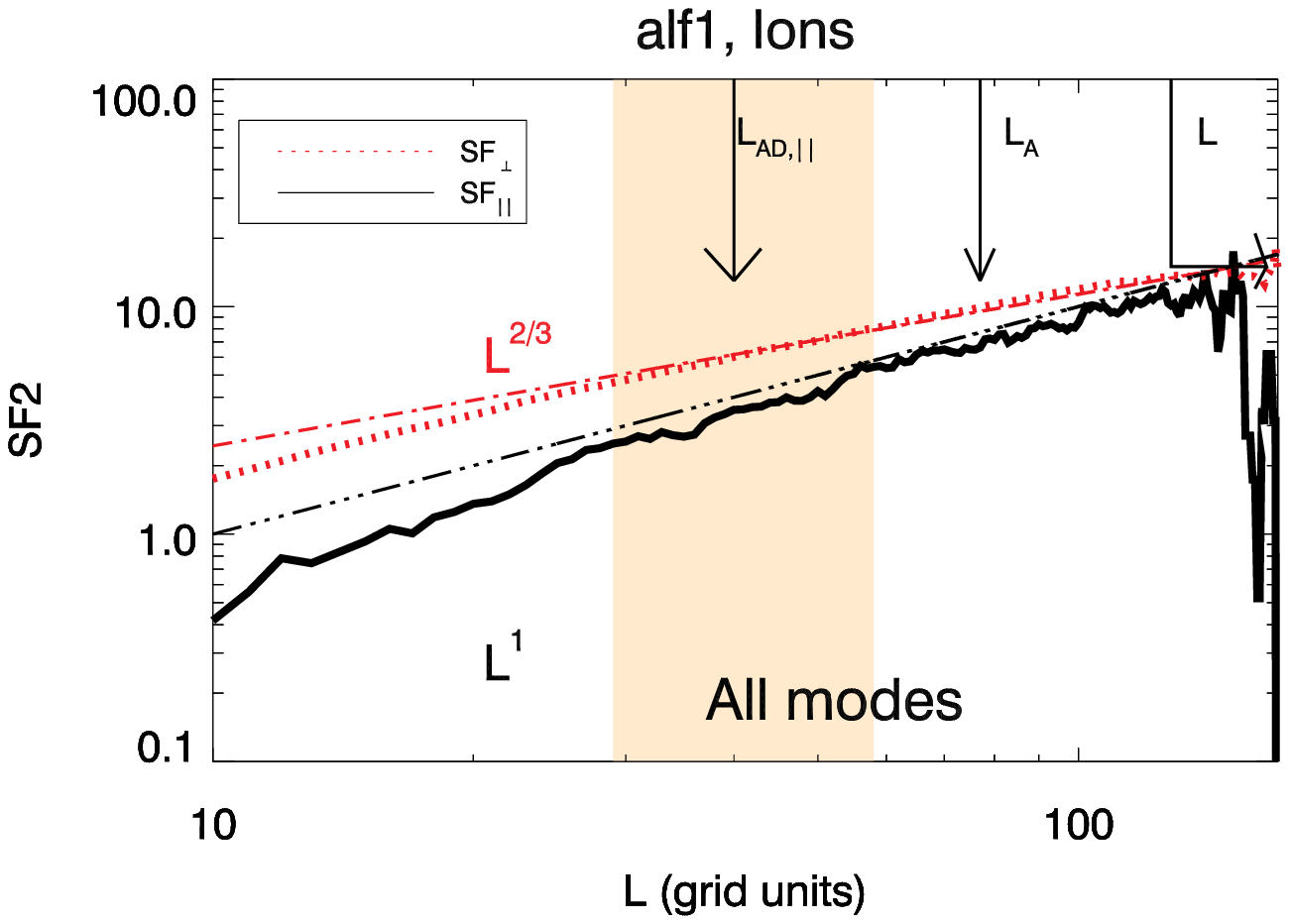}
\includegraphics[scale=.5]{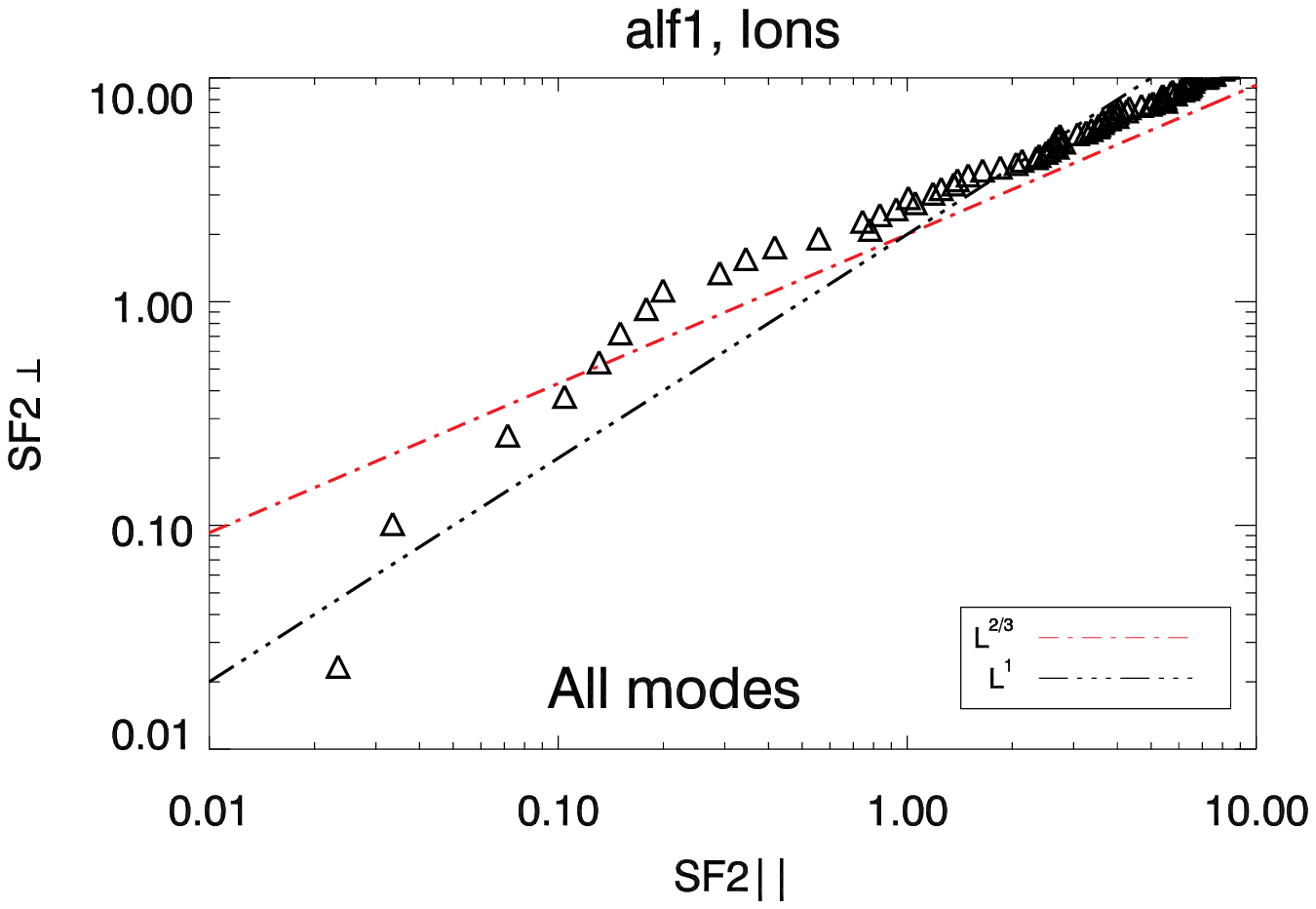}

\includegraphics[scale=.5]{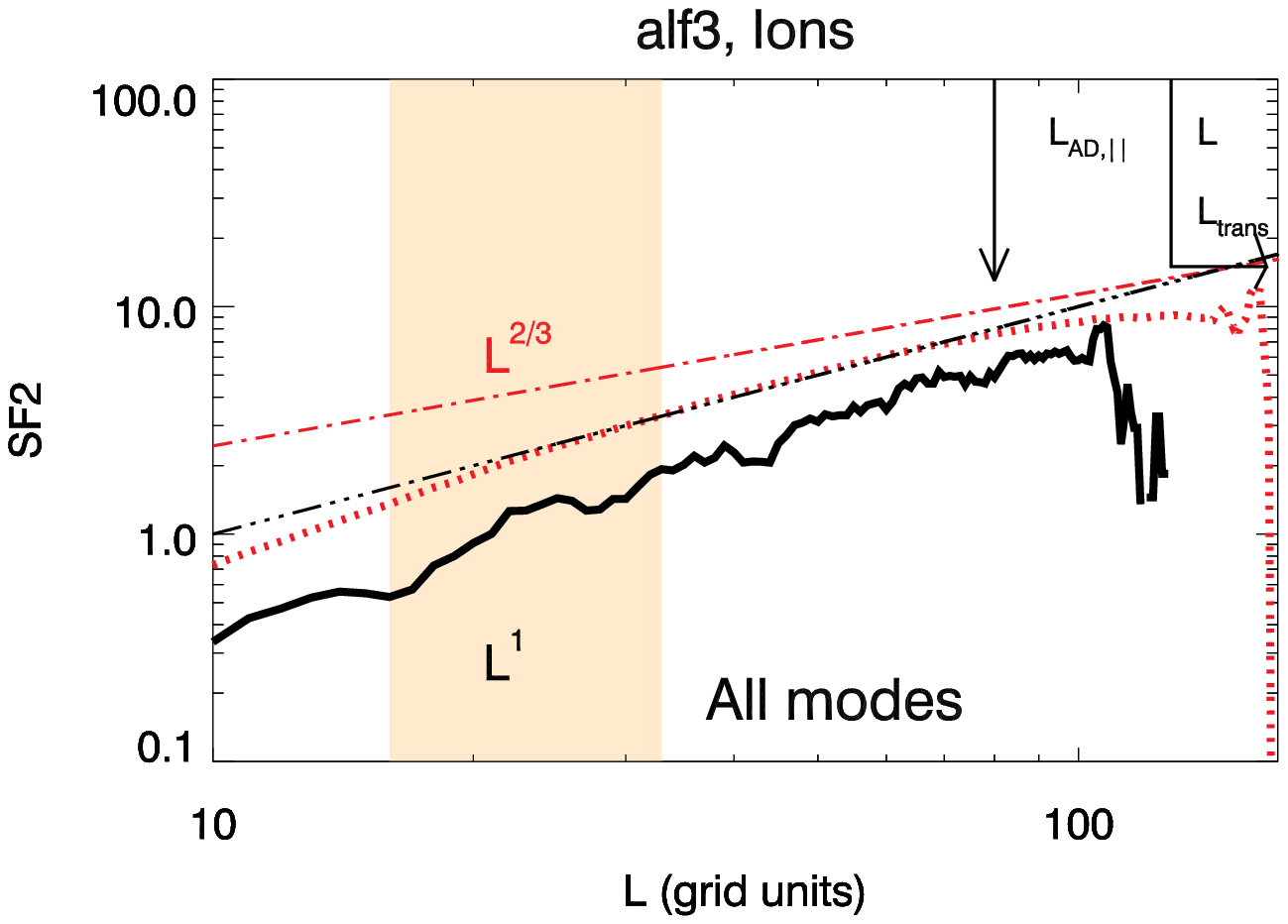}
\includegraphics[scale=.5]{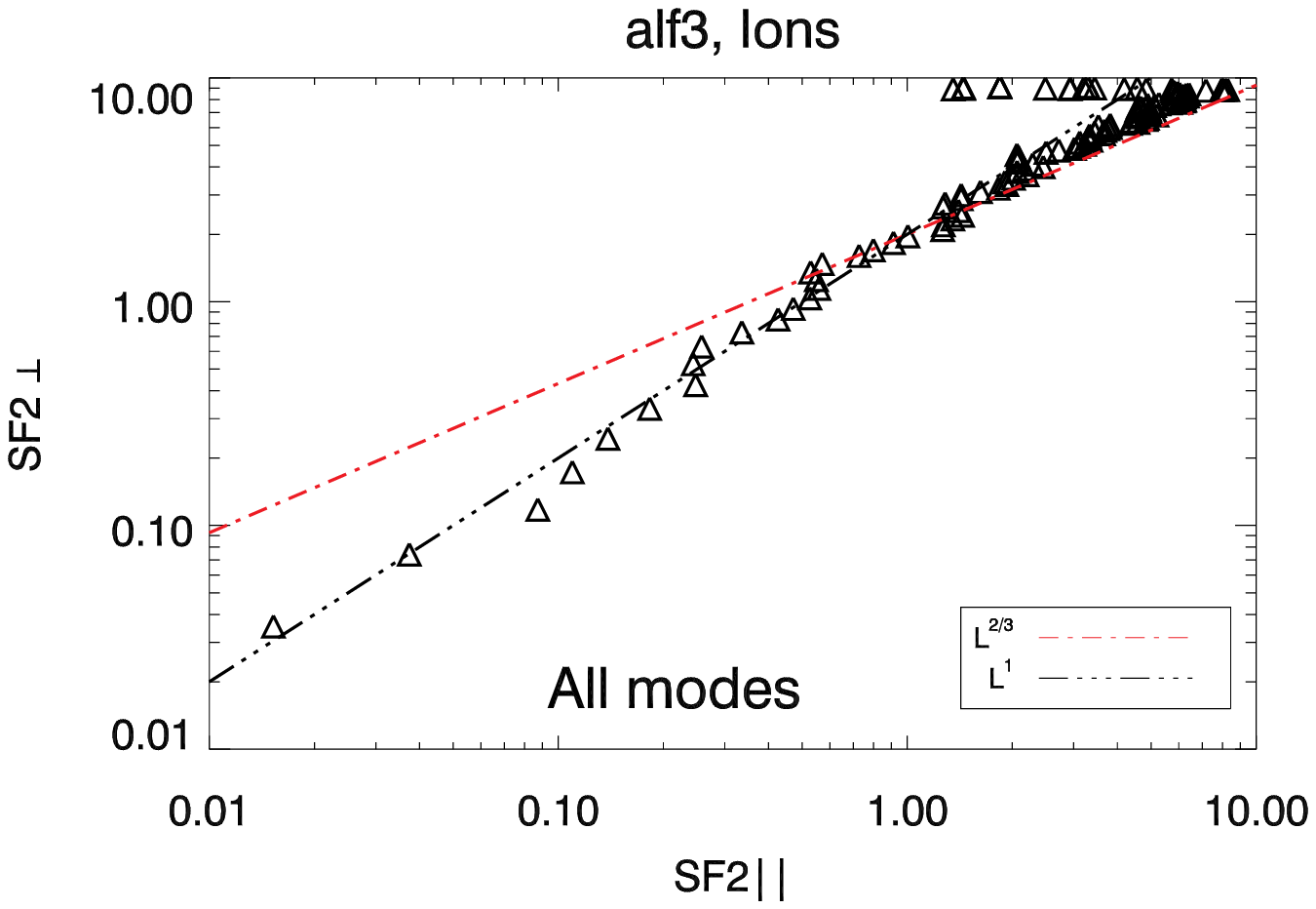}

\includegraphics[scale=.5]{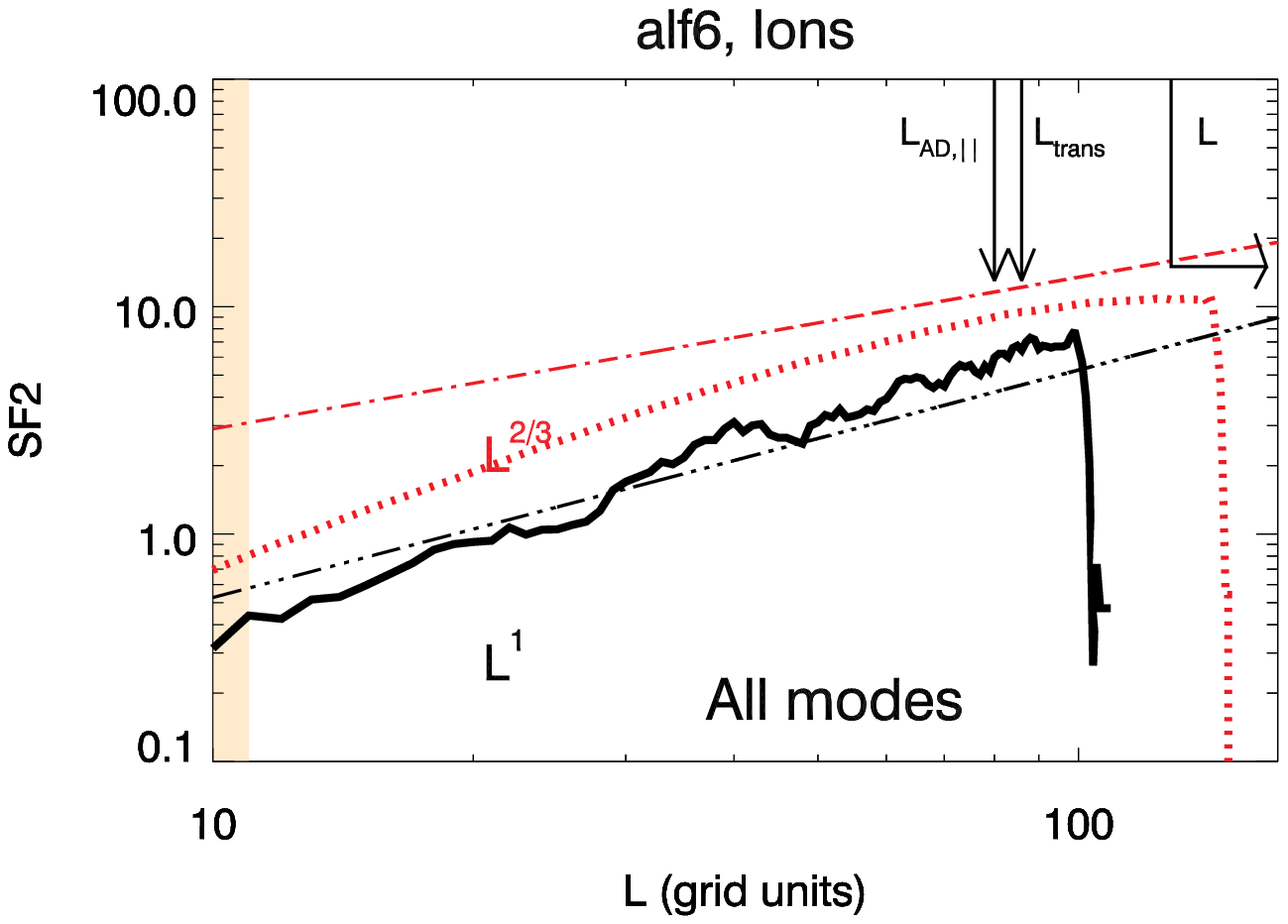}
\includegraphics[scale=.5]{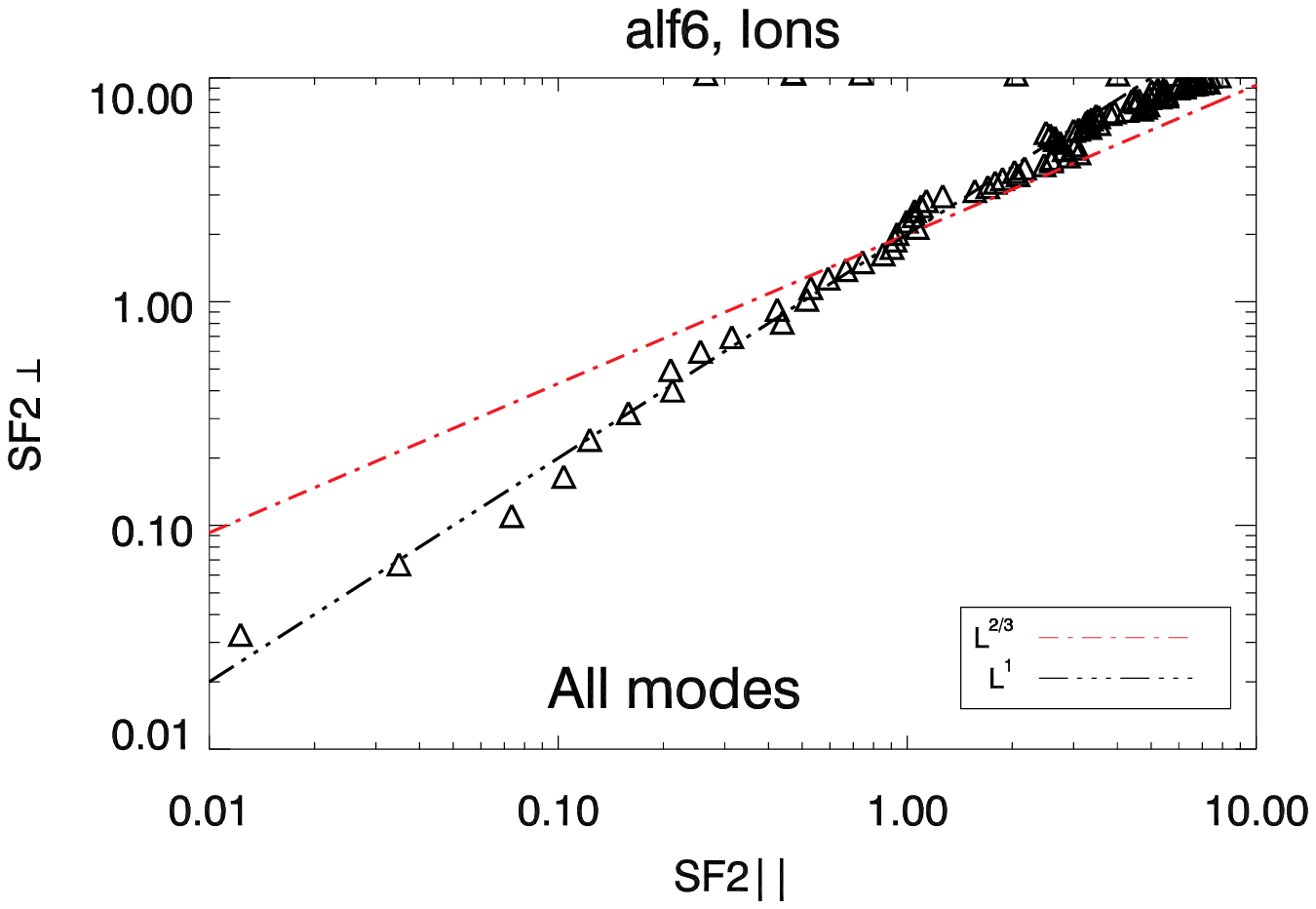}
\caption{Structure functions  of the ion velocity field in the local frame of reference.  
 The two columns both show the structure function and their predicted scalings (red color lines for the perpendicular and black color lines for
the parallel scaling).  The first column shows both parallel and perpendicular structure functions vs. scale while the second column plots the parallel
vs. perpendicular structure functions. Row one present the results for model A1 (which is super-Alfv\'enic) and row two and three present
results for models A3 and A6 (sub-Alfv\'enic), respectively.  In the left column, we indicate the presence of several important scales: the driving scale (L), the parallel ambipolar diffusion
scale (L$_{AD, \|}$), the perpendicular ambipolar diffusion scale (range of tan box using driving scales from 512-128 pixels), 
and the transition scale to GS95 (L$_A$ for model A1 and L$_{trans}$ for models 
A3 and A6).
\label{fig:allion}}
\end{figure*}

\begin{figure*}[tbhp]
\centering

\includegraphics[scale=.5]{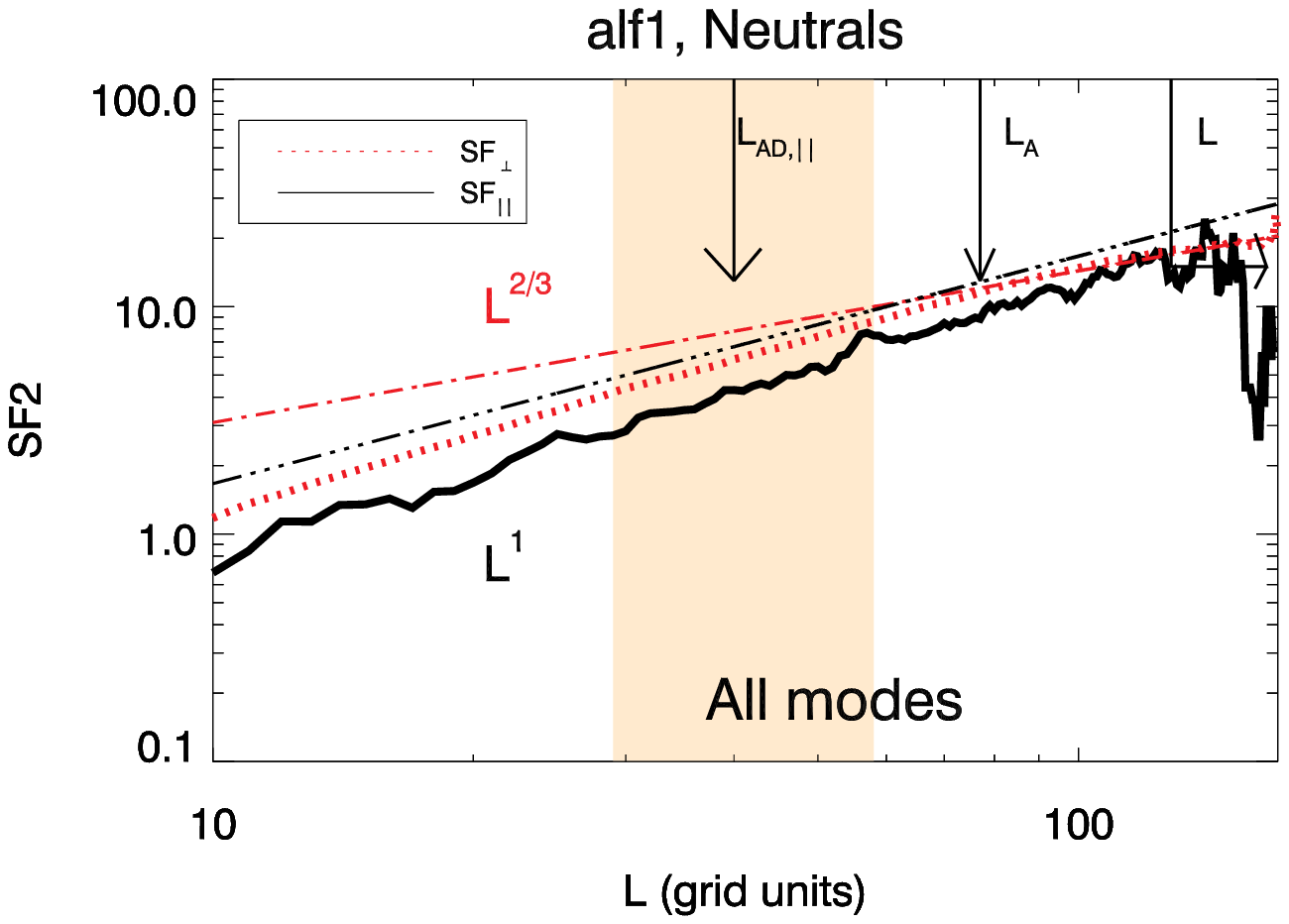}
\includegraphics[scale=.5]{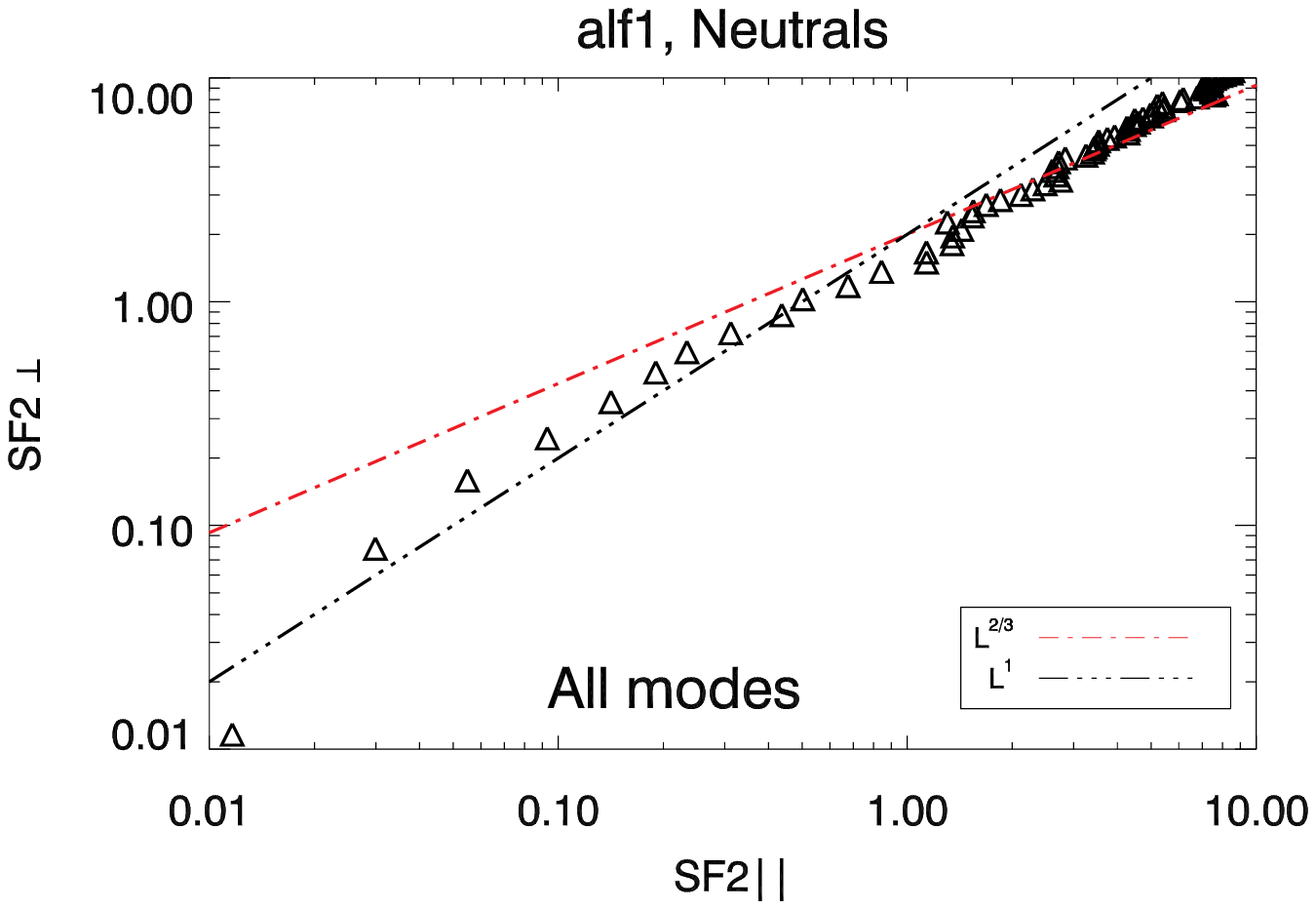}

\includegraphics[scale=.5]{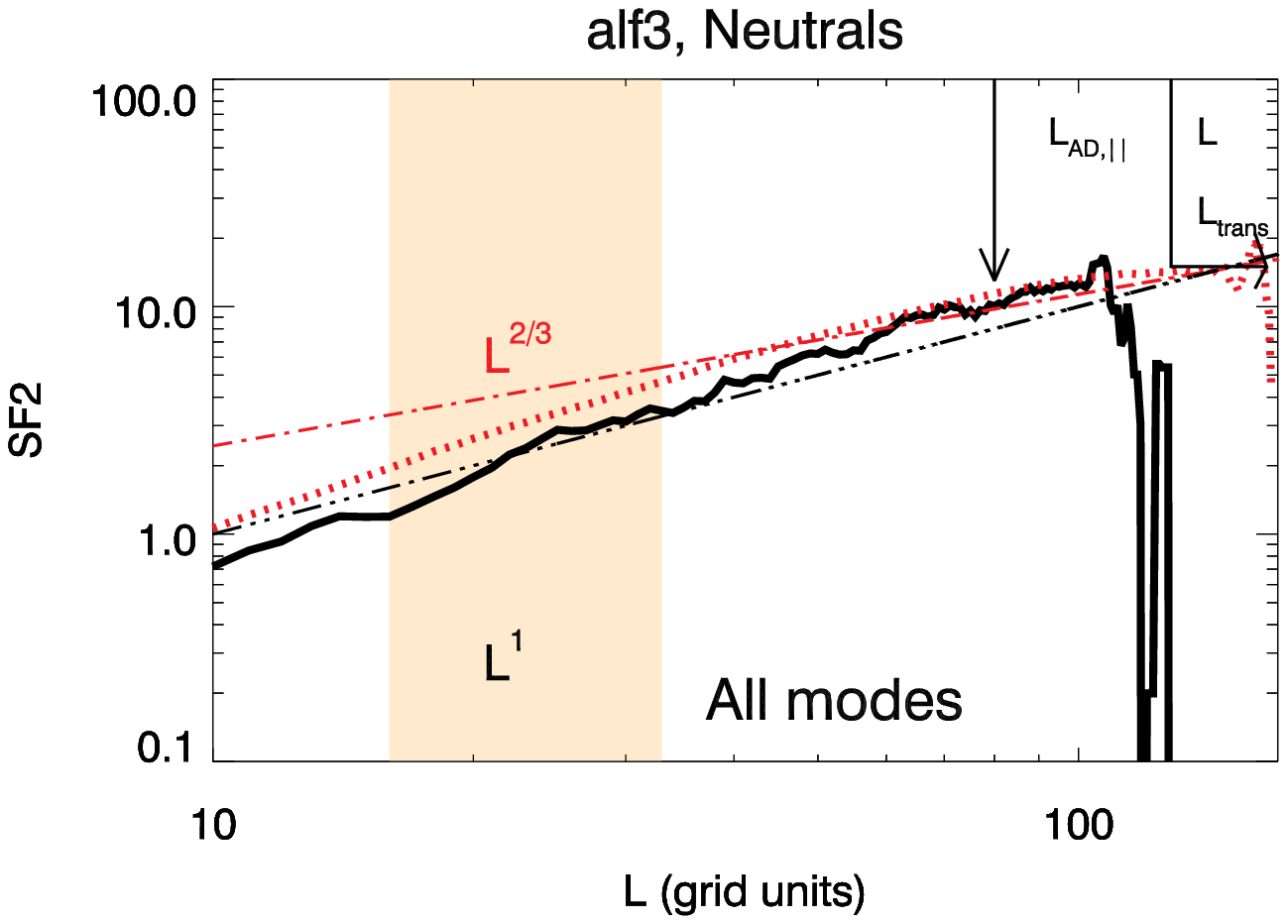}
\includegraphics[scale=.5]{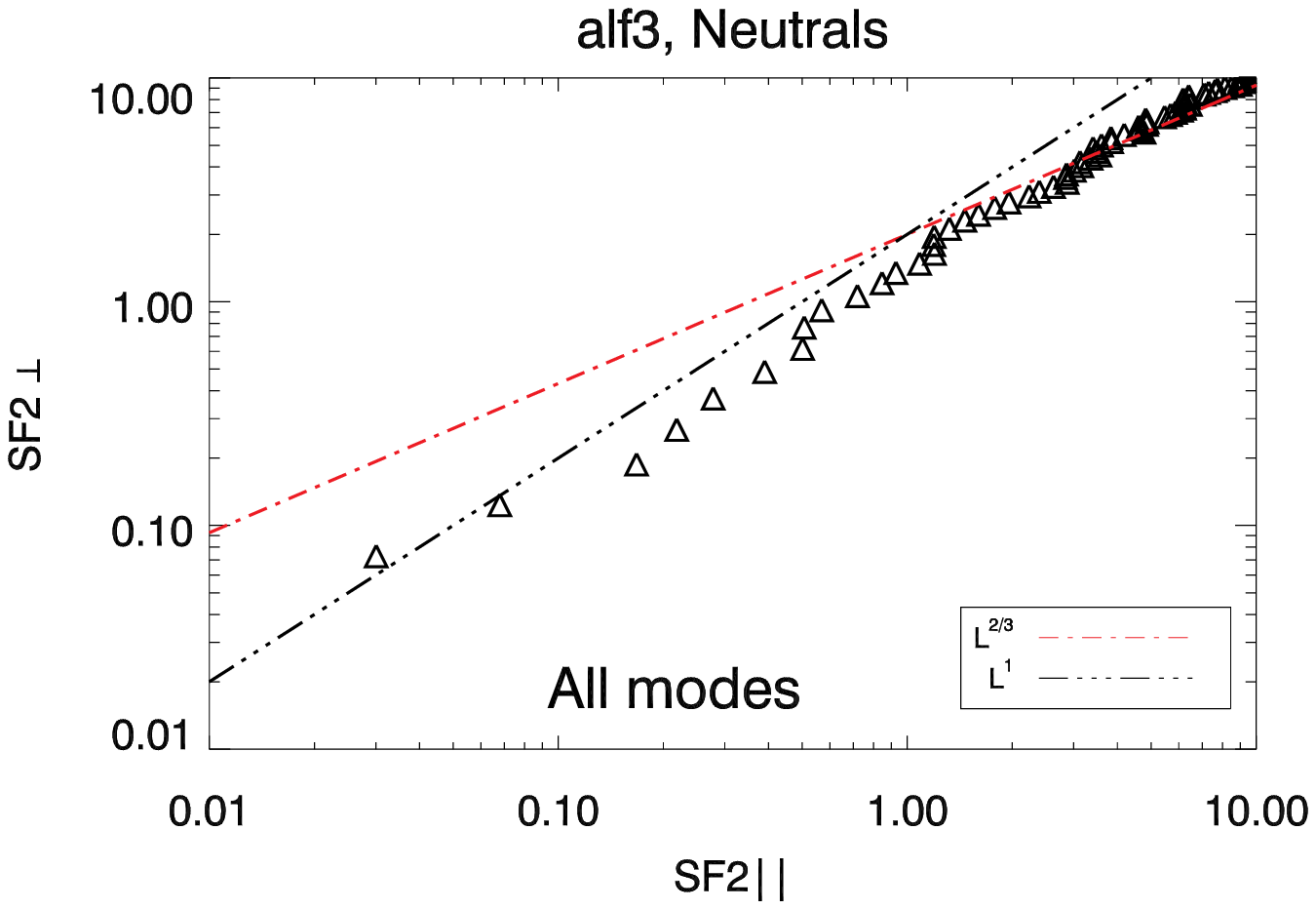}

\includegraphics[scale=.5]{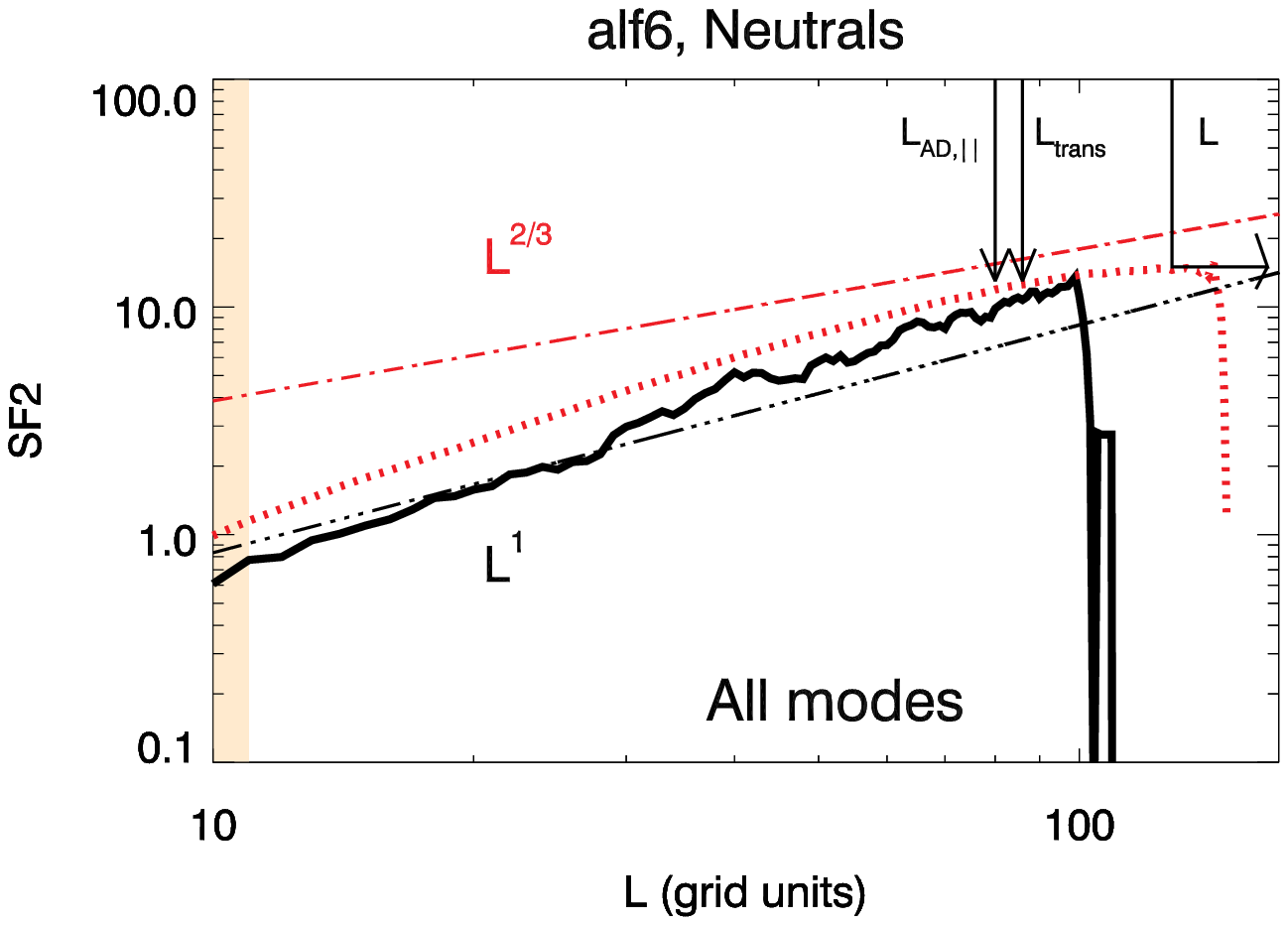}
\includegraphics[scale=.5]{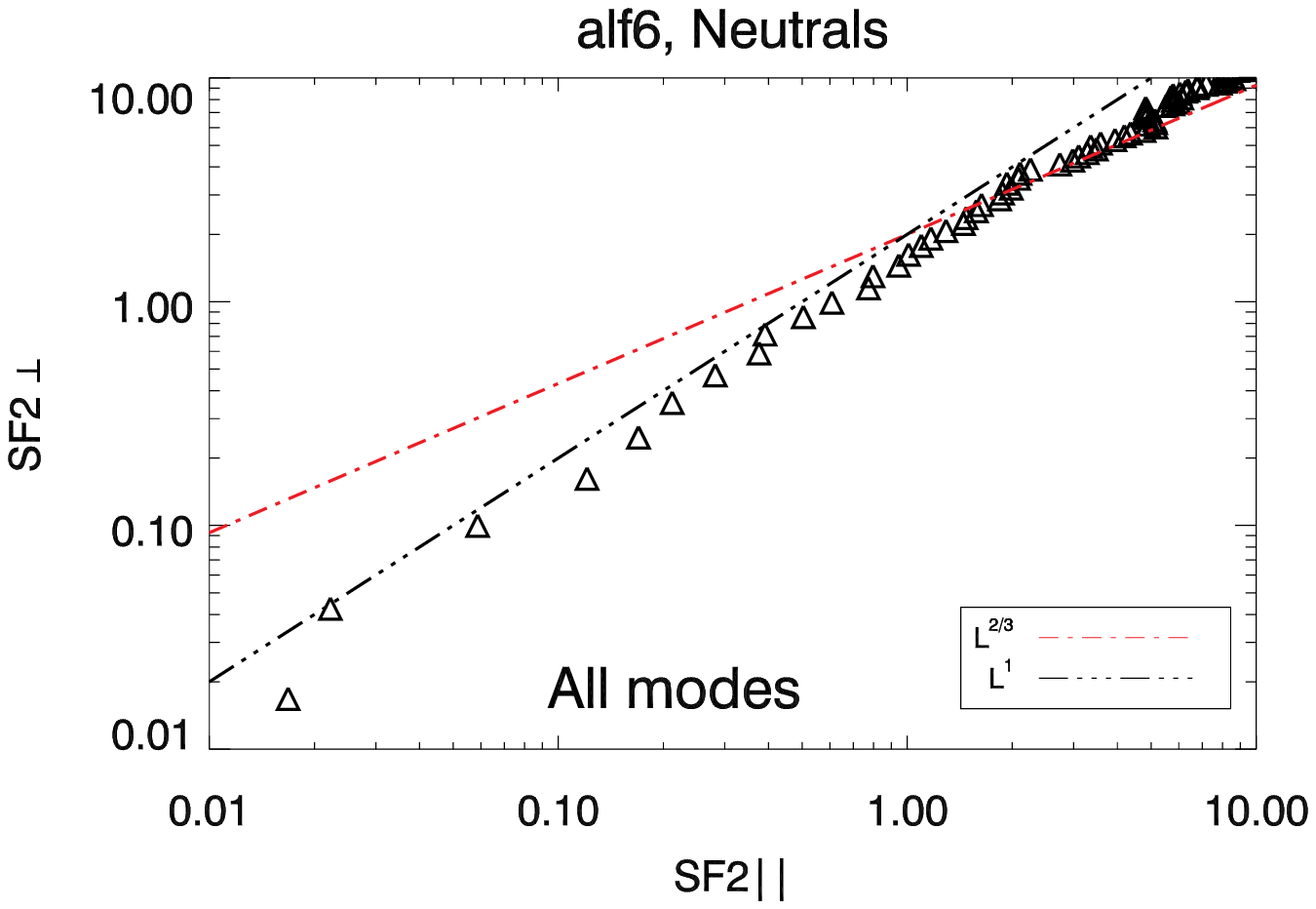}
\caption{Structure functions  of the neutral velocity field in the local frame of reference.  
The figure is organized in an identical manner to Figure 1).
\label{fig:allneu}}
\end{figure*}

We perform mode decomposition of the velocity field of the ions and neutrals as described in Cho \& Lazarian (2003).
We briefly summarize this procedure and direct the reader to the original work by Cho \& Lazarian (2003) for a more detailed presentation.
The slow and fast velocity components can be obtained by projecting the velocity Fourier component $v_k$ onto $\xi_s$ and $\xi_f$, 
where $\xi_f$ and $\xi_s$ are the basis vectors for the slow and fast modes, respectively, which lie in the plane defined by the mean magnetic field 
direction $\textbf{B}_0$ and $\textbf{k}$.  They are defined as:
\begin{equation}
\hat{\xi_s} \propto (-1 +\alpha_a -\sqrt{D})k_{\parallel}\bf{\hat{k_{\parallel}}}+(1+\alpha-\sqrt{D})k_{\perp}\hat{k}_{\perp}
\end{equation}

\begin{equation}
 \hat{\xi}_f \propto (-1 +\alpha_a +\sqrt{D})k_{\parallel}\bf{\hat{k_{\parallel}}}+(1+\alpha+\sqrt{D})k_{\perp}\hat{k}_{\perp}
\end{equation}

and the basis vector for the Alfv\'en mode is:
\begin{equation}
\hat{\xi_A} =\hat{\bf{k_{\perp}}}\times\hat{\bf{k_{\parallel}}}
\end{equation}

where $D=(1+\alpha_a)^2-4\alpha_a cos(\theta)$, $\alpha_a=c_s^2/V_A^2=\beta(\gamma/2)$.
$\theta$ is the angle between $\textbf{k}$ and $\textbf{B}_0$, $c_s$ is the sound speed and $\gamma$ is the adiabatic index. 
 These expressions for the mode decomposition basis are rigorously 
derived in Appendix A of Cho \& Lazarian (2003) and illustrated in their Figure 2.

\section{Results}
\label{nomode}
\subsection{Velocity scaling without mode decomposition}

First we investigate the structure functions of the velocity field in the local frame relative to the mean magnetic field.  In this subsection we present our results  {\it without} mode decomposition.
We calculate the structure functions:
\begin{equation}
SF_2(\bf{r})=<|v(\bf{x+r})-v(\bf{x})|^2>
\end{equation} 
in which we obtain separately the cases in which the axis is aligned parallel and perpendicular with the \textit{local} mean field

We  present the structure function analysis for the ion velocity field in Figure \ref{fig:allion} and for the neutral velocity field in Figure \ref{fig:allneu}.
In the ions for all simulations,  the GS95 scaling is observed till L$_{AD, \|}$ and there is a range of
scales in which the scaling of $l^{2/3}$ can be seen for all models in the right column panels.
Model A1 (top panels) is super-Alfv\'enic and thus the transition to strong turbulence and the GS95 scaling does not begin until scale L$_A\approx77$ grid units however past this scale a scaling of $ l^{2/3}$ is clearly seen down to L$_{AD, \|}$

Models A3 and A6 (middle and bottom panels, respectively), are sub-Alfv\'enic and thus are in the regime of weak or wave-like turbulence until
scale $L_{trans}$ given in Table 1.    Both A3 and A6 have  L$_{AD, \|}$ and  L$_{AD, \bot}$ separated by a larger dynamic range
of scales than model A1 due to sub-Alfv\'enic turbulence.  At scales smaller than the driving scale and $L_{trans}$, Alfv\'enic turbulence in the ions develops and is clearly seen in the left column panels by the scaling relation $l_{\|} \sim l^{2/3}$. 
For models A3 and A6, the range of scales for GS95 is limited (as seen in the right column panels), as either the driving scale or L$_{trans}$  is close to  L$_{AD, \|}$.
However for model A1 the range of L$_{AD}$ is separated from both L and L$_A$ with sufficient dynamic range to generate an Alfv\'enic cascade with the
GS95 scaling in the ions.  

 The neutral structure functions shown in Figure \ref{fig:allneu}  also have GS95 scalings until  L$_{AD, \|}$ in all three simulations studied. At $L=L_{AD, \|}$ neutrals decouple from the ions and no longer exhibit the scaling relations of an MHD cascade. The $l^{2/3}$ scaling is present in the neutrals until L=40 for model A1.  Model A3 and A6 transition to an $l^{1}$ scaling at larger scales than model A1 with model A6 transitioning to $l^{1}$ at slightly larger scales then model A3. Since models A3 and A6 have identical values for L$_{AD, \|}$ this suggests that the driving scale and/or Alfv\'en Mach number also plays a role in the dissipation scale of partially ionized fluid turbulence.
It is also interesting to note that in the A1 model the ions exhibit a greater dynamic range then the neutrals over which the $l^{2/3}$ is observed.  This suggests that the ions may continue the MHD cascade even after they have decoupled from the neutrals.  As most of the energy in the MHD cascade lies in the Alfv\'en modes (see Kowal \& Lazarian 2010) it is important to determine if the damping of the cascade at these scales is due to damping in the Alfv\'en modes.

\subsection{Velocity scaling with mode decomposition}

In order to address the nature of the Alf\'enic cascade in ion-neutral turbulence we must separate the Alfv\'en modes from the Fast and Slow modes.  We perform the mode decomposition on the the ion and neutral velocity fields to extract the Alfv\'en modes and then apply the second order structure function analysis as was shown in Figures \ref{fig:allion}  and \ref{fig:allneu}.

\begin{figure*}[tbhp]
\centering

\includegraphics[scale=.5]{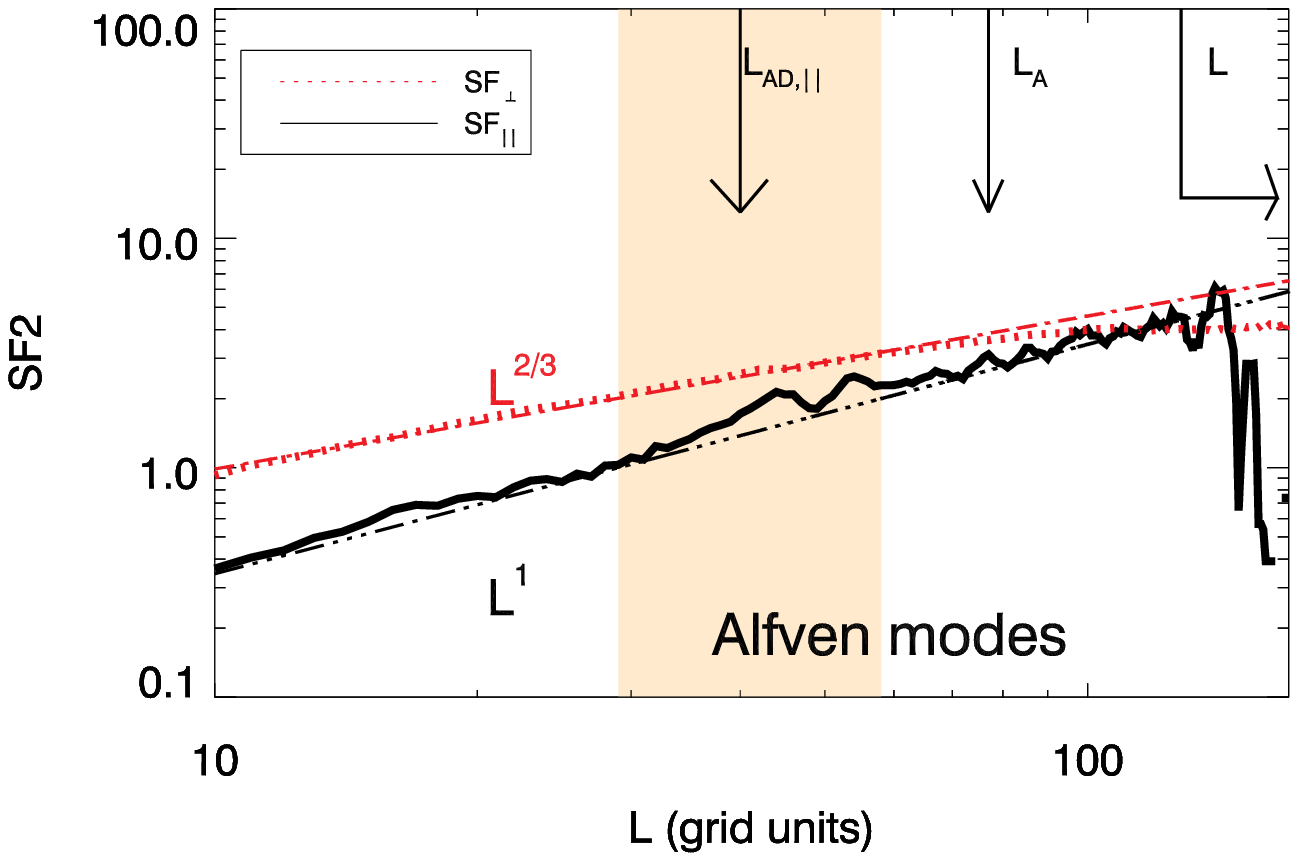}
\includegraphics[scale=.5]{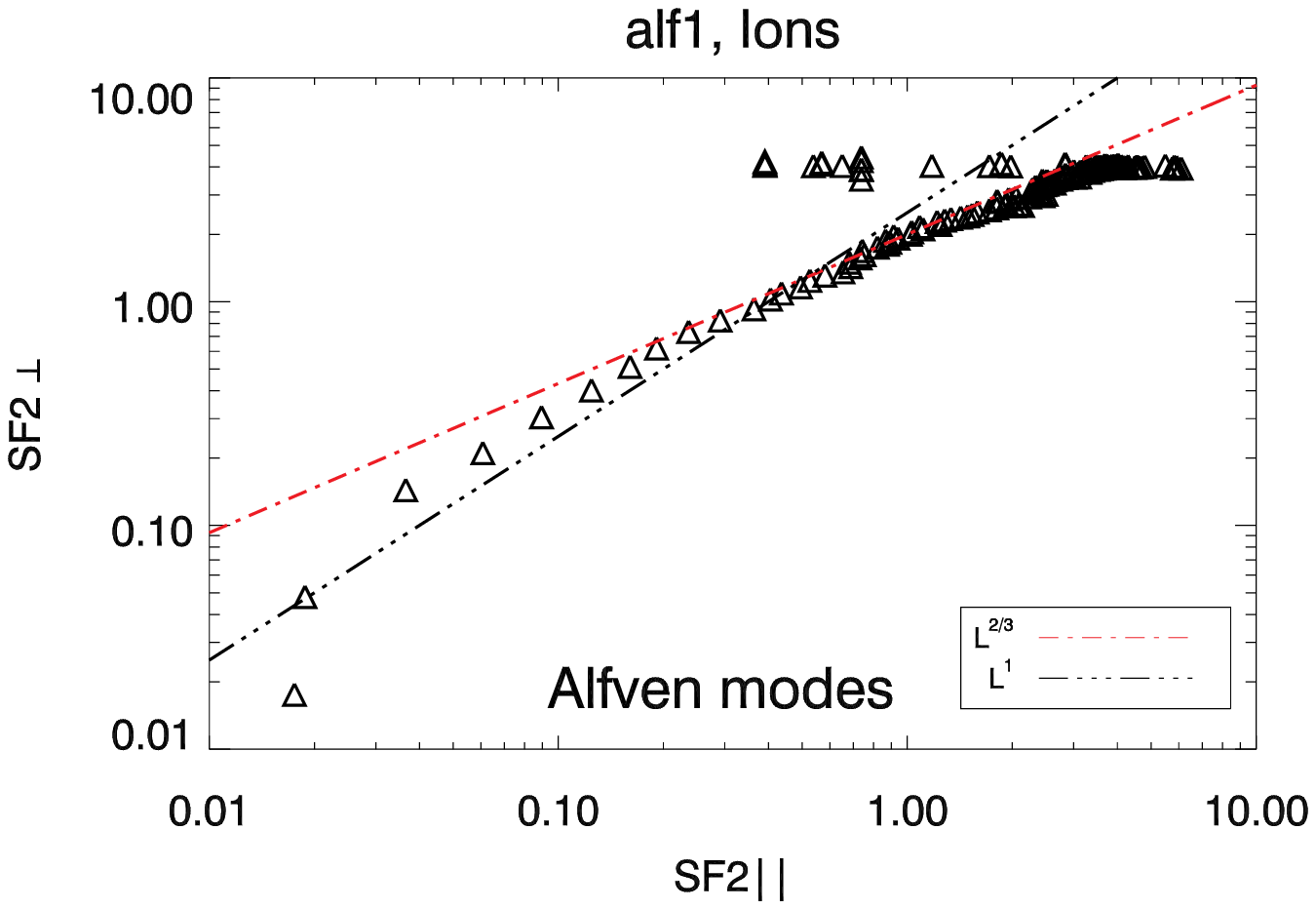}

\includegraphics[scale=.5]{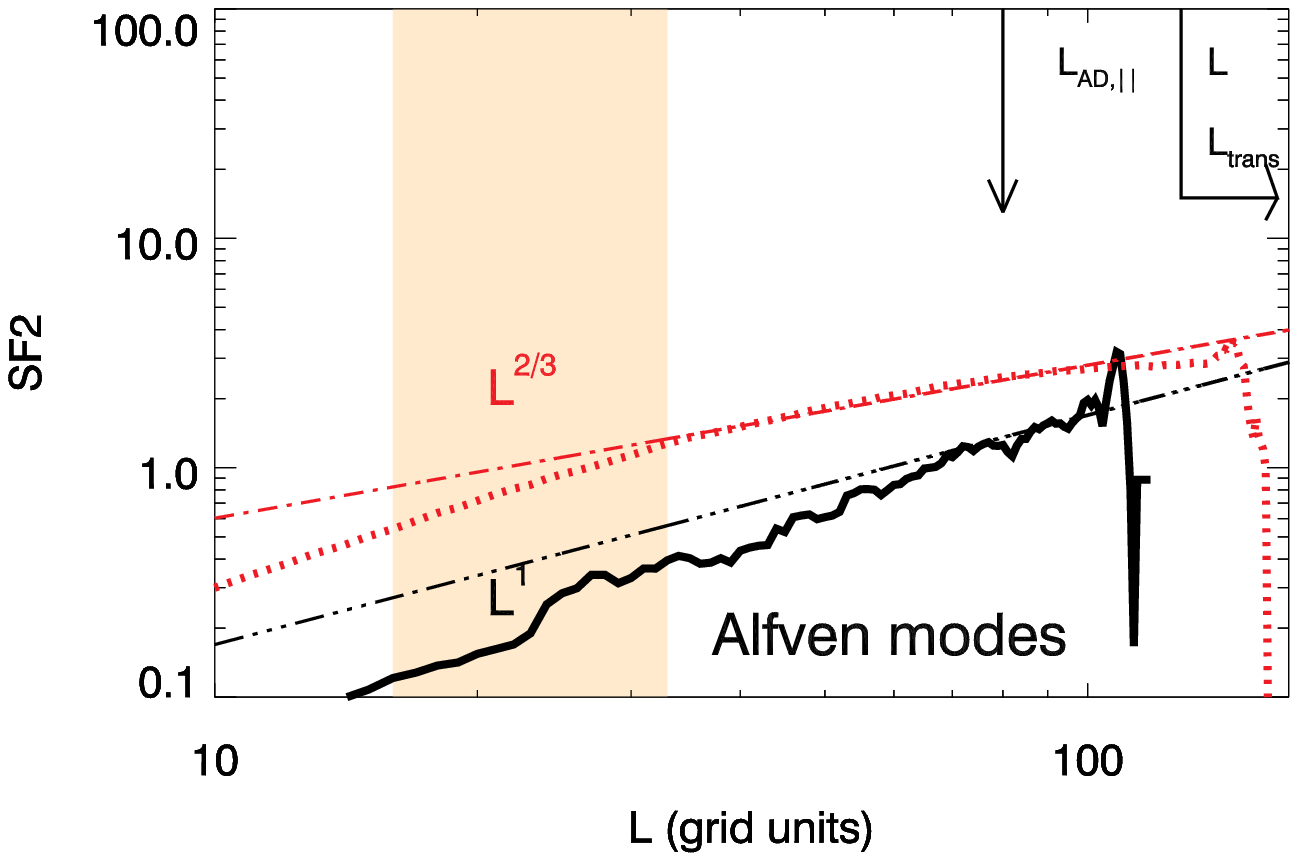}
\includegraphics[scale=.5]{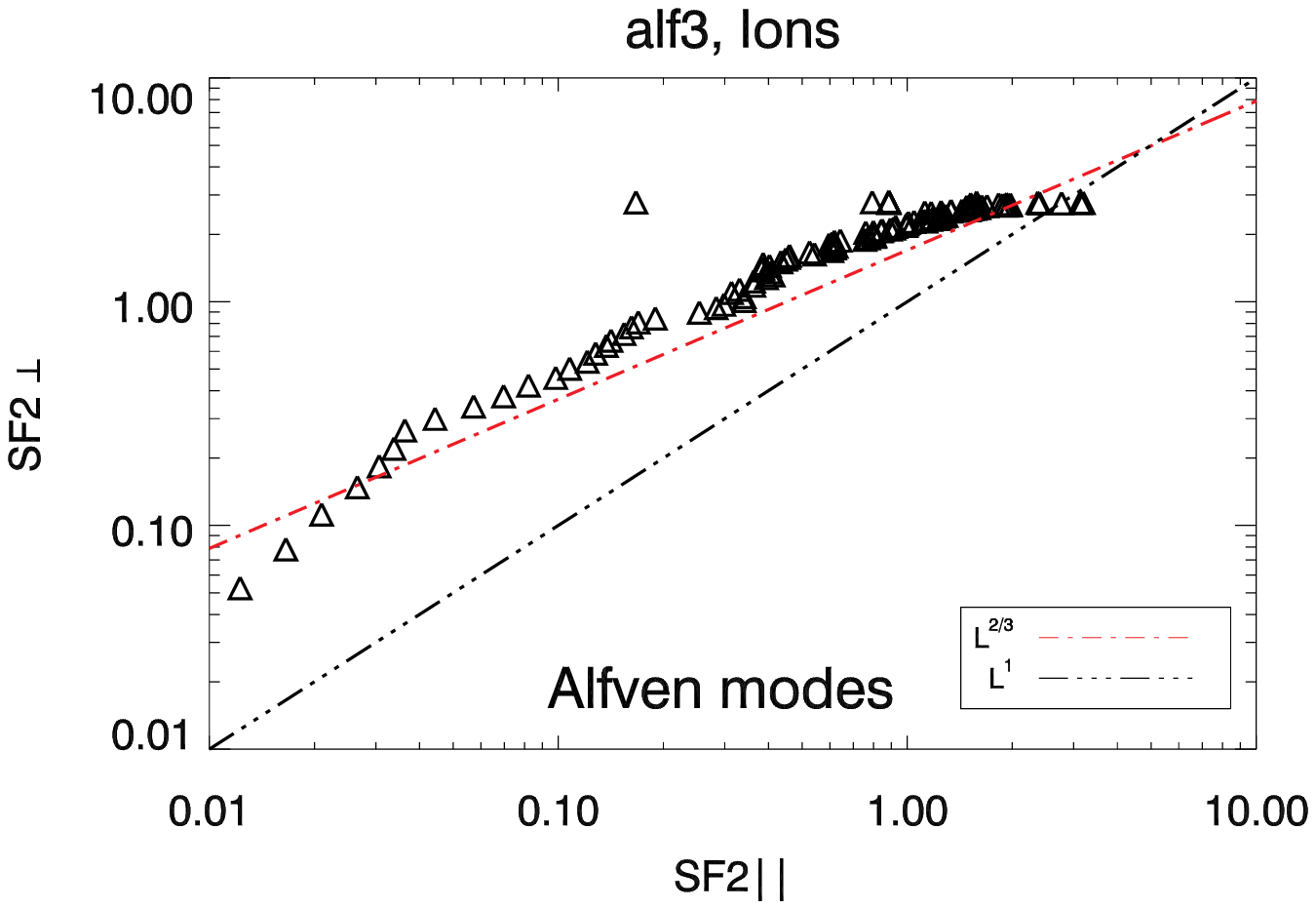}

\includegraphics[scale=.5]{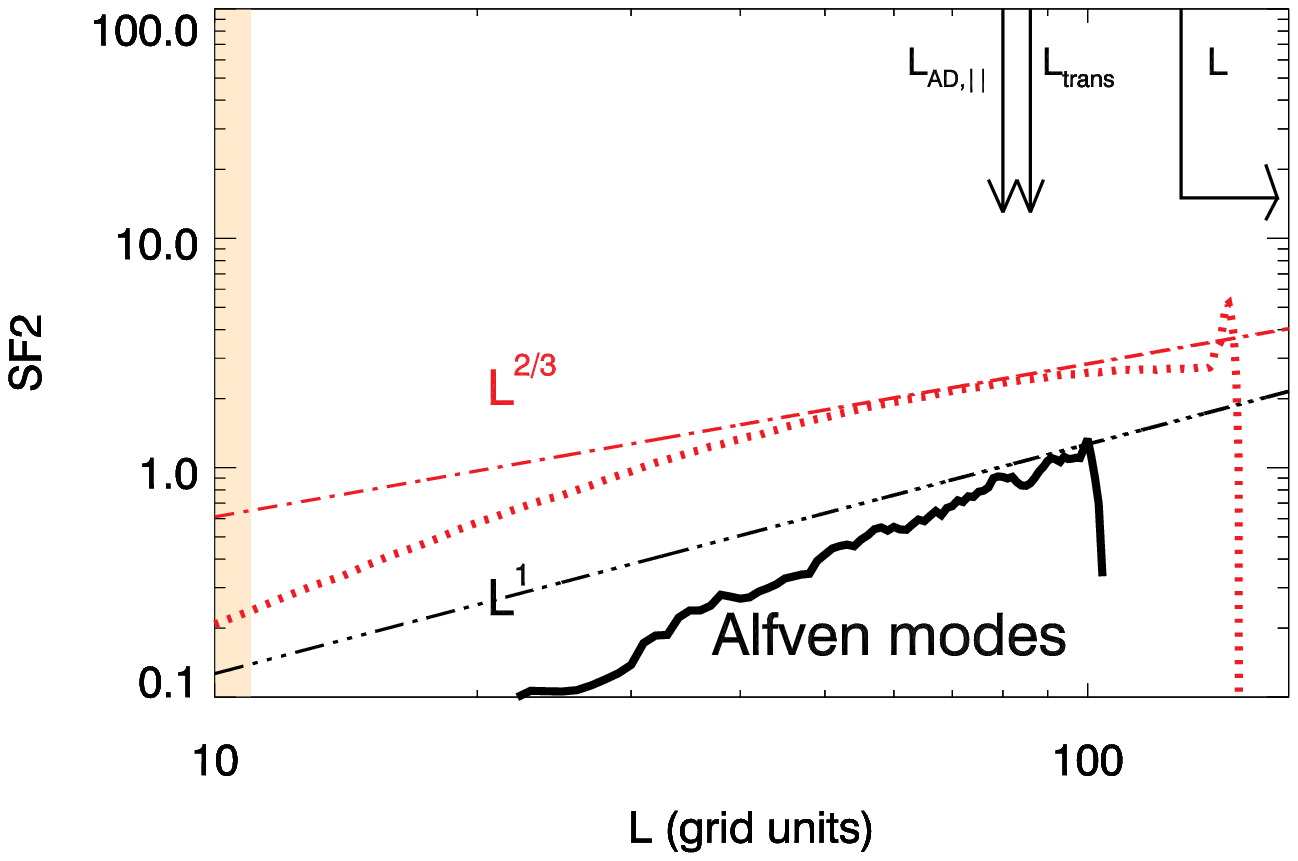}
\includegraphics[scale=.5]{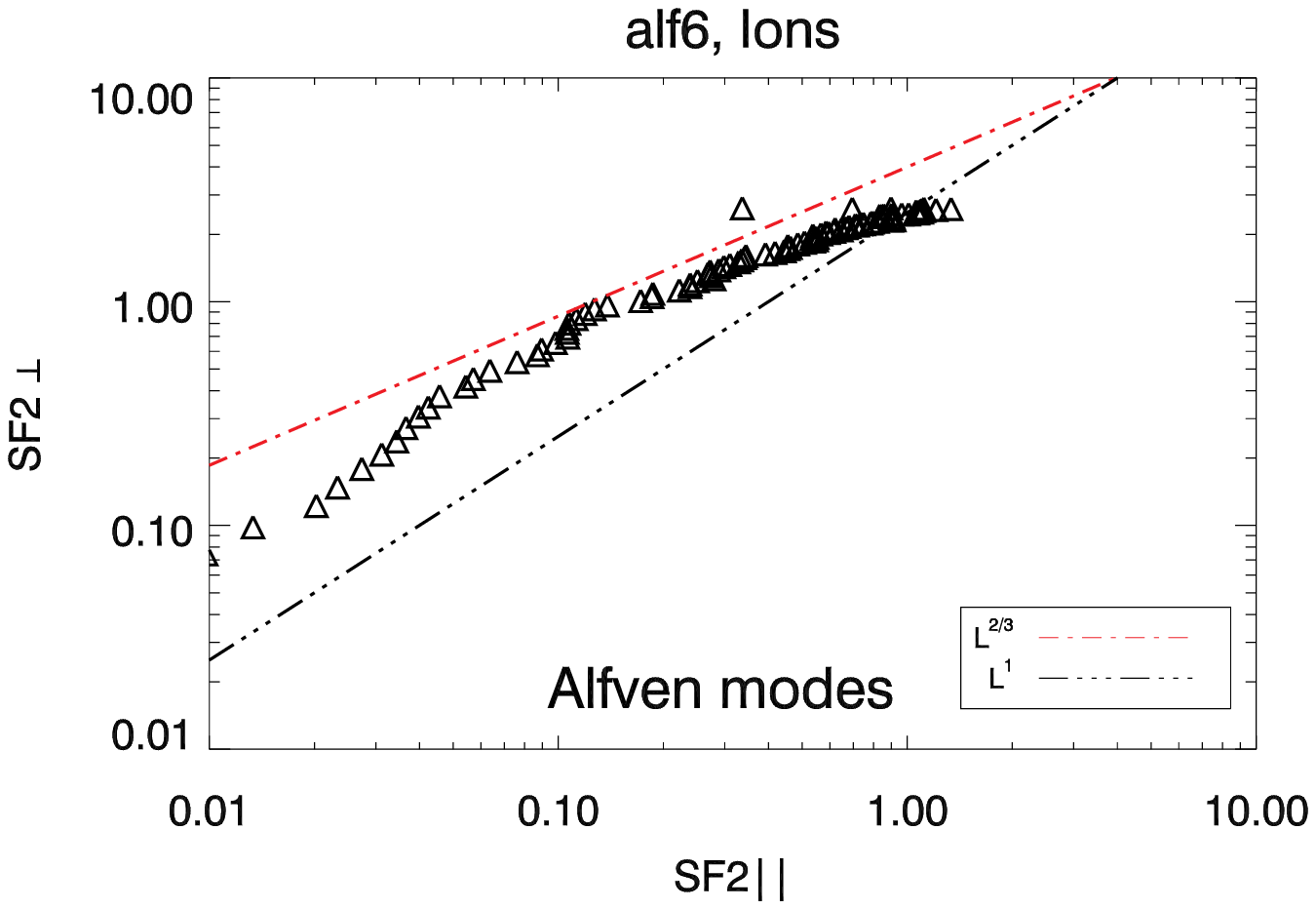}
\caption{Structure functions for the Alfv\'en modes of the ion velocity field.   The figure is organized in an identical manner to
Figure  \ref{fig:allion}.}
\label{fig:alf_sti}
\end{figure*}

We show the ion Alfv\'en mode structure functions in Figure \ref{fig:alf_sti}.
The Alfv\'en modes of model A1 (top left panel)  obey the GS95 scalings to scales smaller than the ambipolar diffusion scale which indicates that the Alfv\'enic cascade can continue to scales smaller than L$_{AD}$.  This suggests that turbulence does not necessarily damp at the ambipolar diffusion scale but can continue to the viscous dissipation scale and would not necessarily provide a characteristic scale for star formation, in agreement with Oishi \& Mac Low (2006).   

Unlike model A1, Alfv\'enic turbulence in the ions begins to damp at larger scales for model A3 and 
A6. Alfv\'enic turbulence in models A3 and A6 damps at or before the perpendicular ambipolar diffusion scale  The larger dynamic range of scales between  L$_{AD, \|}$ and  L$_{AD, \bot}$  coupled with the limited dynamic range 
between the driving scale and  L$_{AD, \|}$,  increases the effect of damping due to ion-neutral
collision.   In addition, the strength of the damping due to ion-neutral collisions goes as $V_A^2$ (Klessen et al. 2000) and explains
why the damping is the strongest for model A6.

\begin{figure*}[tbhp]
\centering

\includegraphics[scale=.5]{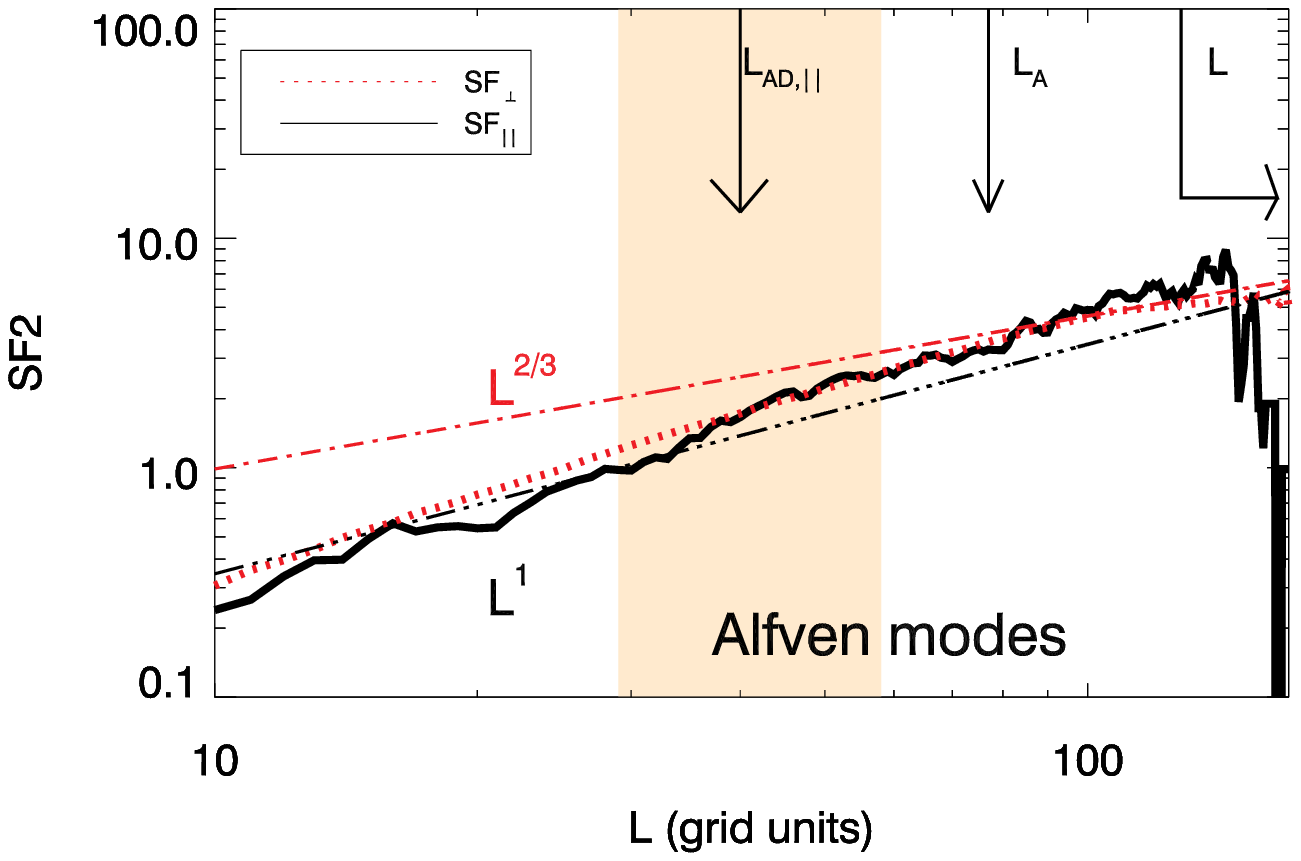}
\includegraphics[scale=.5]{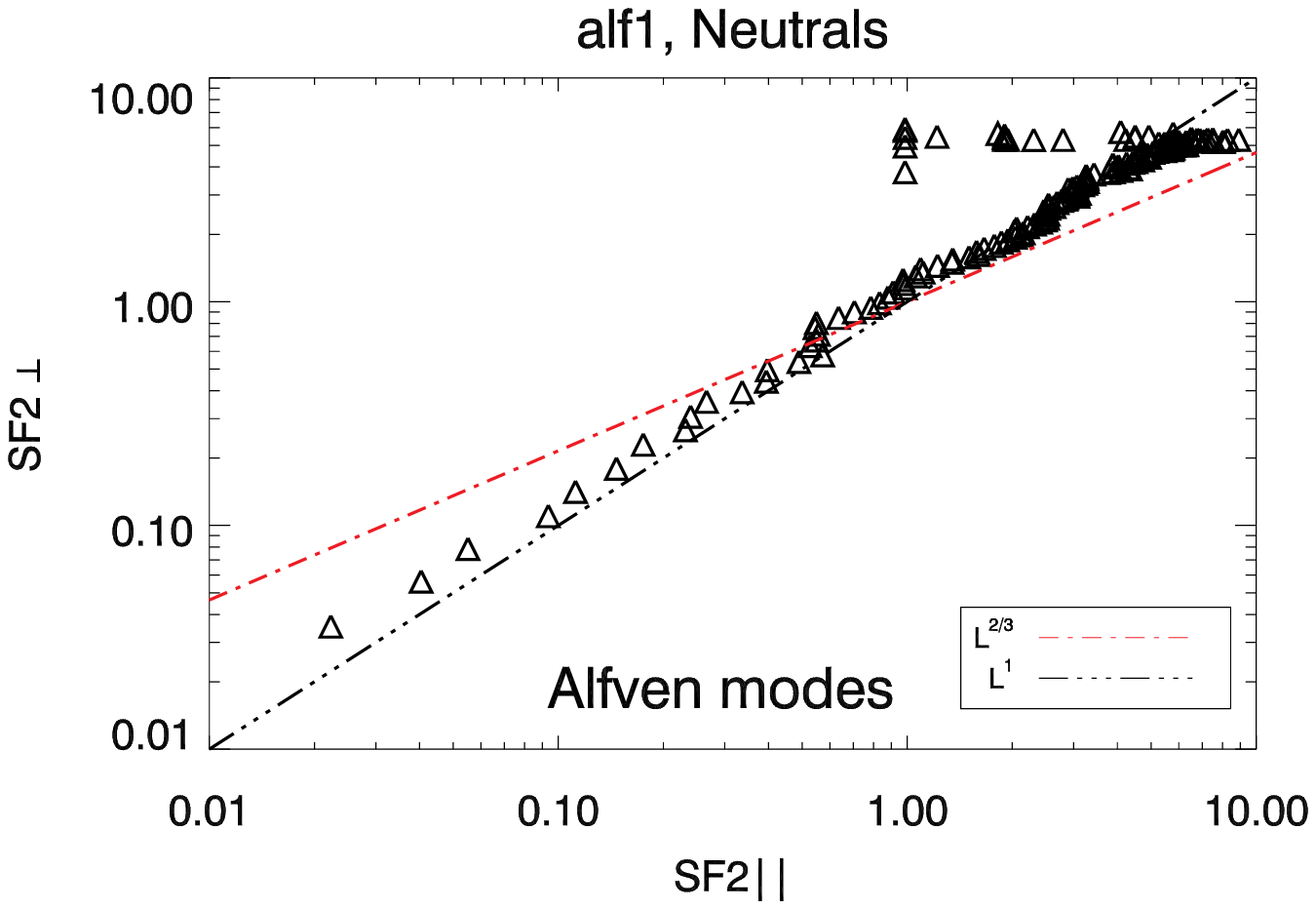}

\includegraphics[scale=.5]{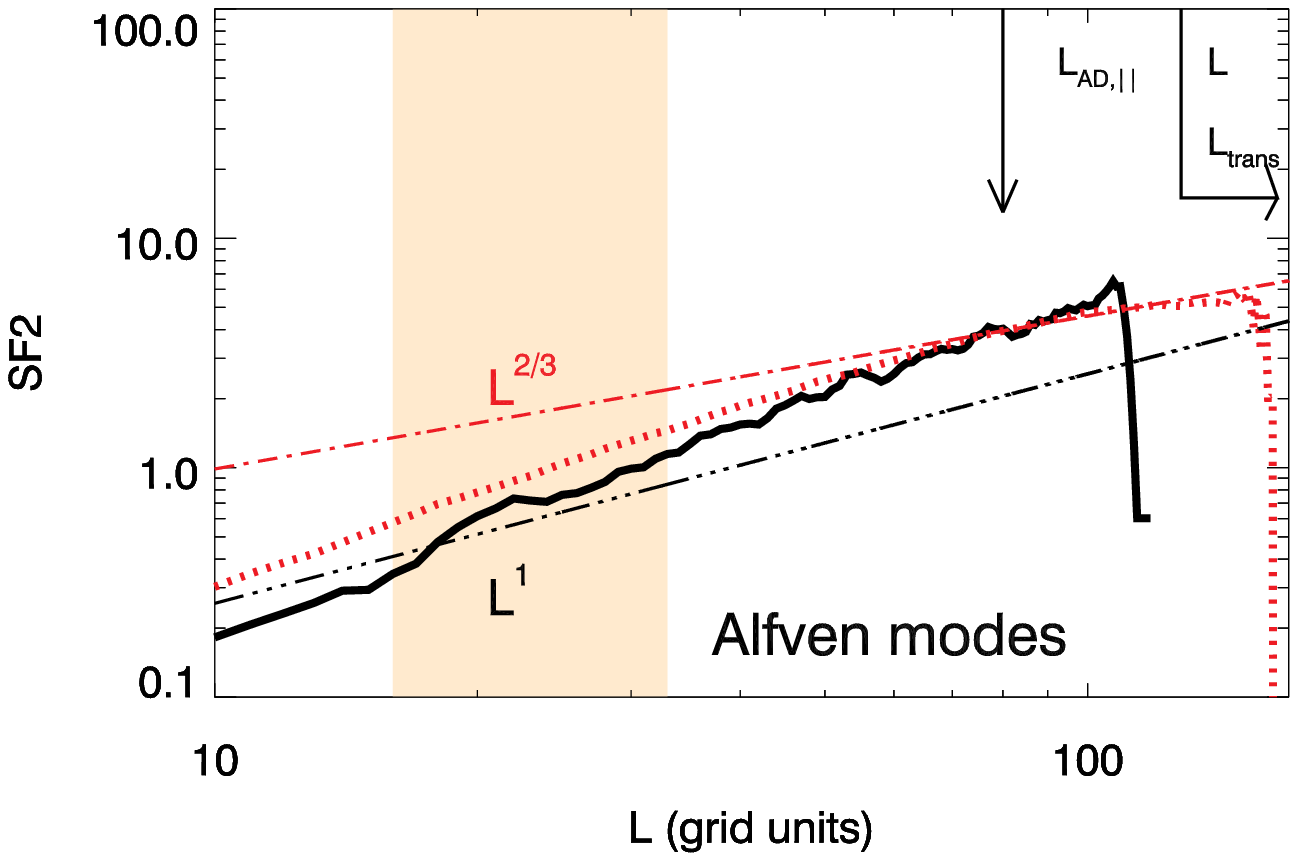}
\includegraphics[scale=.5]{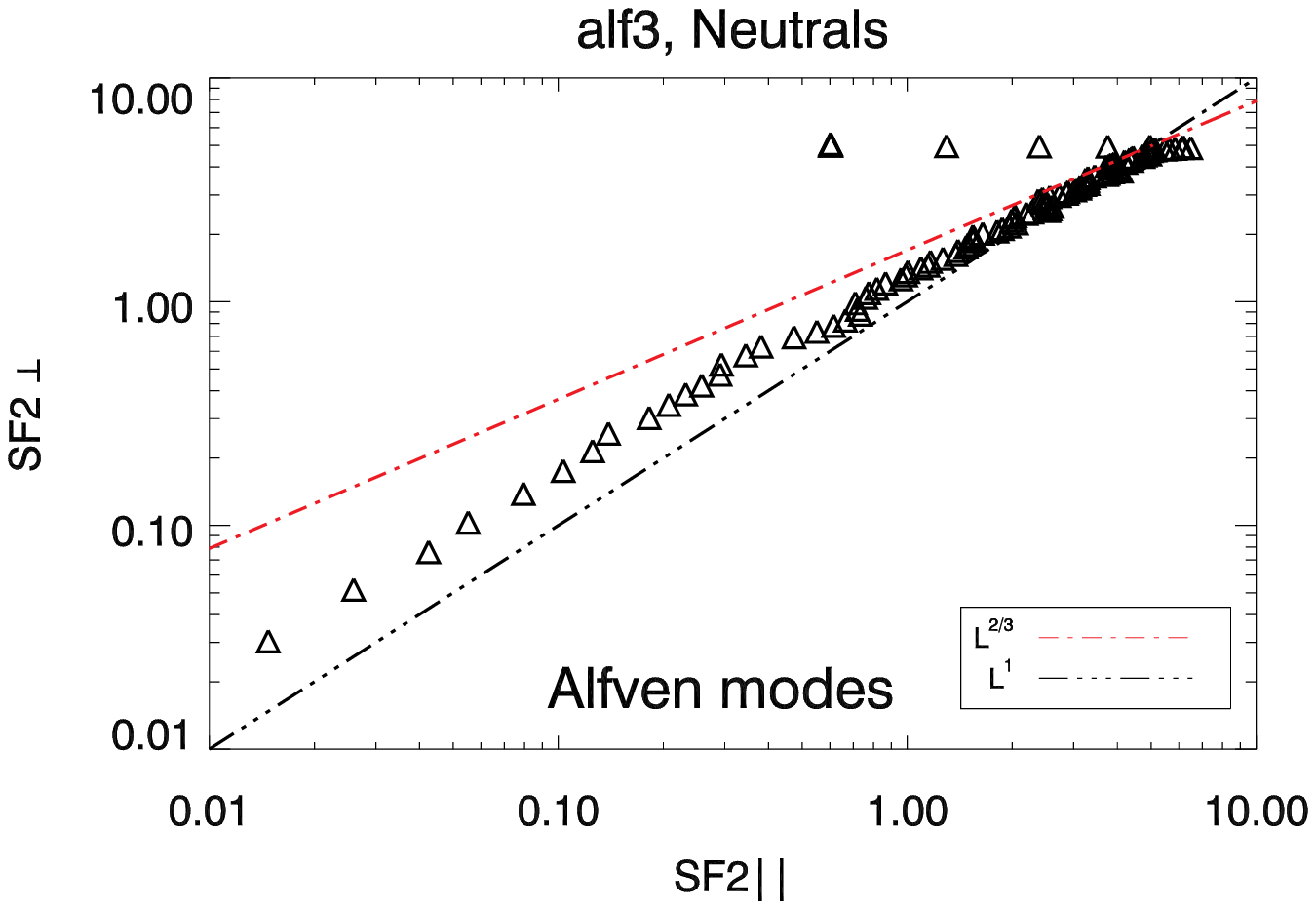}

\includegraphics[scale=.5]{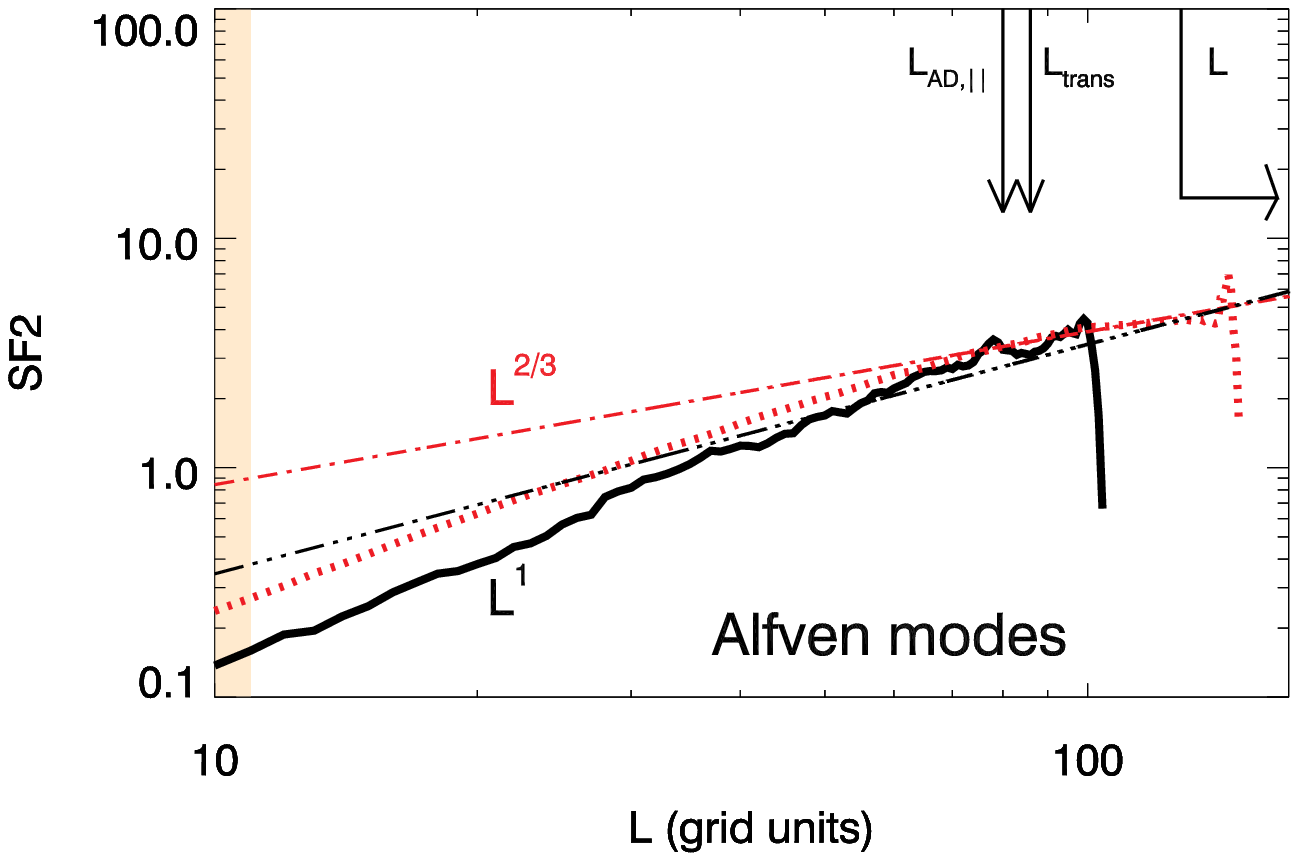}
\includegraphics[scale=.5]{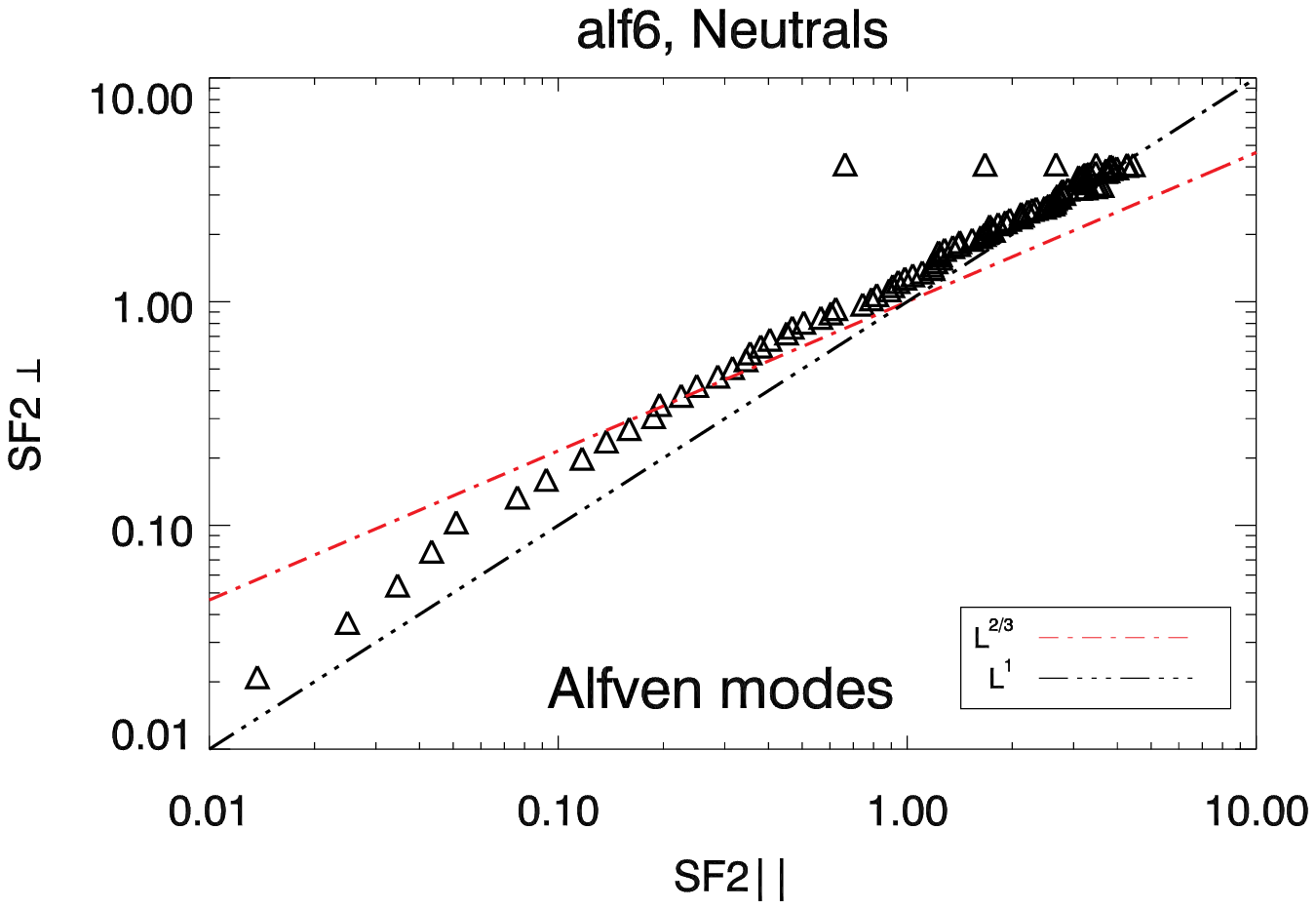}
\caption{Structure functions for the Alfv\'en modes of the neutral velocity field. The figure is organized in an identical manner to
Figure  \ref{fig:allion}. Once the neutrals have decoupled from the ions the mode decomposition to separate the Alfv\'en modes from the fast and slow modes
is no longer a physically motivated procedure.}
\label{fig:alf_stn}
\end{figure*}

Figure \ref{fig:alf_stn} plots the structure functions 
of the Alfv\'en mode neutral velocity field of each model in Table 1. The organization of the figure is identical to Figure \ref{fig:alf_sti}. 

In all simulations, the Alfv\'en mode scaling in the neutrals are seen at large scales but break off from the GS95 scaling at scales smaller than $L_{AD,\|}$ and/or $L_{AD,\bot}$.
In all three models the perpendicular motions do not follow  $l_{\|} \sim l^{2/3}$ at scales smaller than $L_{AD,\|}$  but another effect seen is that
both  $l_{\|}$ and  $l_{\bot}$ begin to steepen to slopes greater than $l^{1}$.  This is close to the numerical dissipation scale but also most-likely an unphysical effect due to 
the fact that once the neutrals decouple from the ions we can no long apply mode decomposition as Alfv\'en modes will not exist in the neutral cascade.

What are the possible damping mechanisms that can explain the behavior seen in these simulations?
The role of neutral-ion damping in MHD turbulence has
been discussed previously in the context of the ISM (in particular, see Spangler 1991; Minter \& Spangler 1997;LG01; LVC04). 
In a partially ionized medium a combination of
neutral particle viscosity and ion-neutral collisional coupling
drives damping. Neutral-ion friction will compete with and eventually dominate the Alfv\'en wave restoring force
which will damp oscillations in the magnetic field. 
As soon as the neutral-ion collisional rate
is approximately equal to the eddy turnover time, the neutrals will begin to form a hydrodynamic cascade.
 At this scale, and all smaller scales, the ionic fluid
motions will damp at the rate of ion-neutral collisions. 
The damping mechanism in A3/A6 is most likely the result of ion-neutral collisions however  an additional
damping mechanism will be provided by neutral viscosity, which maybe non-negligible in dissipating the MHD cascade in partially ionized gases (LVC04).

\subsection{Power Spectrum}

Finally, we investigate the kinetic energy spectra for simulation with and without mode decomposition.
We calculate the energy power spectra as:  
\begin{equation}
P(\vec{k})=\sum_{\vec{k}=const.}\tilde{A}(\vec{k})\cdot\tilde{A}^{*}(\vec{k})
\end{equation}

In hydrodynamic turbulence the viscous damping scale sets a minimal scale
for turbulent motions and the kinetic energy power spectrum
is dissipated exponentially. This marks the
end of the hydrodynamic cascade, but in MHD turbulence the viscous damping scale
is not the end of magnetic structure evolution.
On these scales magnetic field structures will
be created by shear and magnetic tension. 
As a result, LVC04 predicted a power-law tail in the energy distribution,
rather than an exponential cutoff.

Figure ~\ref{fig:var_lin} shows the neutral kinetic energy power spectrum (black lines)
and the ion kinetic energy power spectrum (red lines) for models A1, A3 and A6.
For model A1, the neutral Alfv\'en modes damp out well before
the ion Alfv\'en modes.  The ions retain the $k^{-5/3}$ 
even after the neutrals have decoupled. 
For models A3 and A6, there is a small portion of the spectrum with the $k^{-5/3}$ scaling
for the Alfv\'en mode in the ions which is not noticeable in the neutrals due to the limited
dynamic range over which the Alfv\'en modes propagate. In the high k damped regime, the slopes approaches
$k^{-4}$ rather than behaving as an exponential decay of energy. This result should confirmed with higher resolution simulations in the future.

\section{Discussion}
\label{discussion}

Supersonic MHD turbulence is known to play a critical role in both the support of GMCs and their subsequent small scale collapse.
Ambipolar diffusion has been thought to be the most important dissipation mechanism for turbulence in GMCs and could set characteristic mass
and length scales for star formation.  However this study shows that turbulence, in particular the Alfv\'enic cascade,
in a partially ionized media does not necessarily damp at the ambipolar diffusion scale L$_{AD}$. 
Other authors, for example Oishi \& Mac Low (2006), investigated two fluid simulations and found that L$_{AD}$
did not set a characteristic length scale  for mass sizes or dissipation and that turbulence may proceed to smaller scales. 
Previous studies, however, often used the heavy ion approximation which dramatically changes the mode propagation characteristics of two fluid turbulence. 
 We show in this work
that Alfv\'enic modes can survive to scales much smaller than L$_{AD}$ in some cases.  This finding also  implies that an important diffusion process called reconnection
diffusion, which is mediated by turbulence (see Lazarian \& Vishniac 1999; LVC04, Lazarian 2005, Santos-Lima et al. 2009, see Lazarian 2014 for a review)  may persist to small scales.
 
The actual damping mechanisms may include neutral particle viscosity and ion-neutral collisions. 
A detailed investigation of the dominance of one damping mechanism over another is beyond 
the scope of this work, however comparisons with Equations 37 and 46 of LG01 suggest that 
the neutral mean free path for our models A3 and A6  is smaller than A1 and hence damping by neutral collisions may play an important role.

\begin{figure*}[]
\centering
\includegraphics[scale=.58]{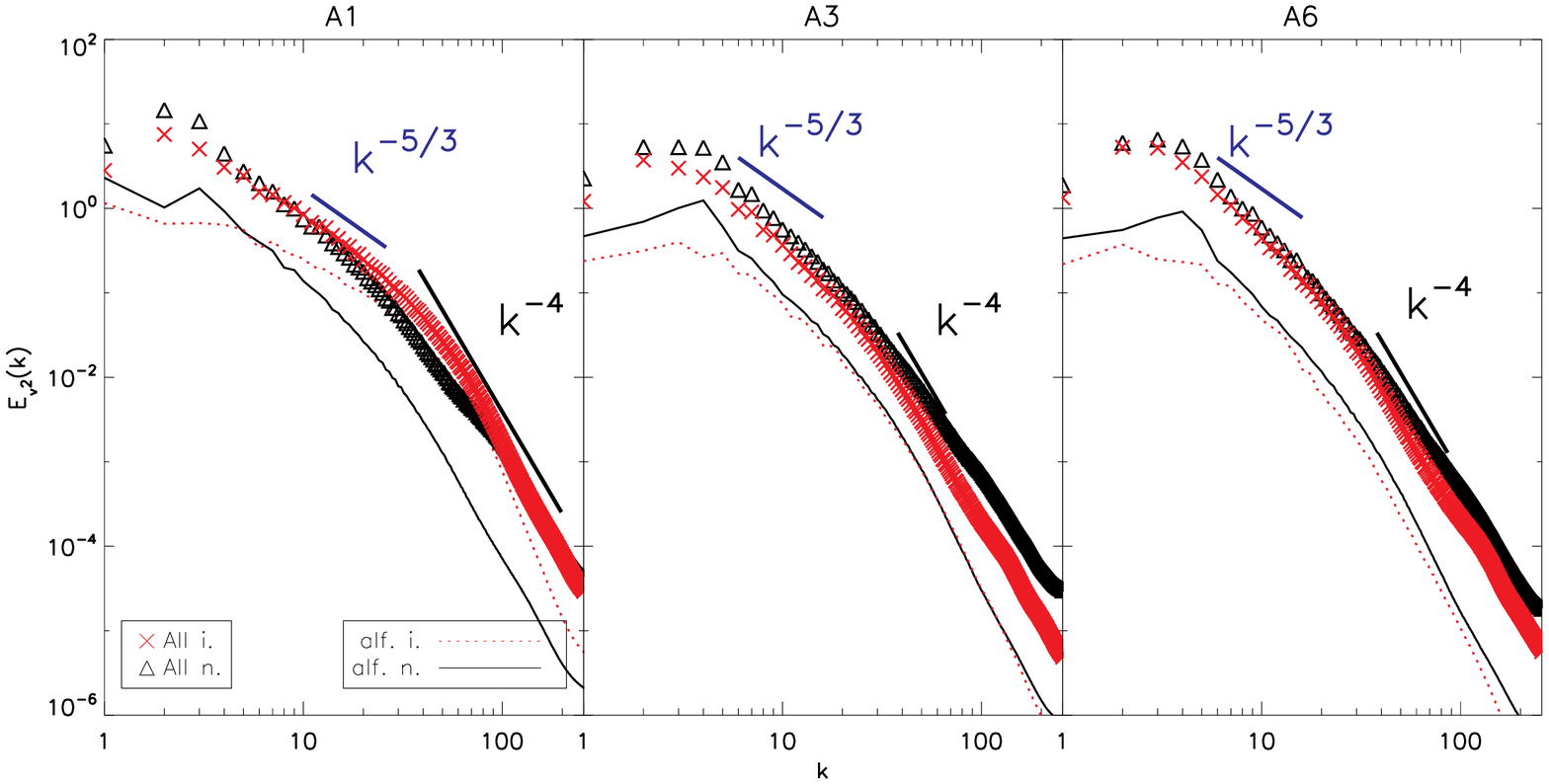}
\caption{The neutral kinetic energy power spectrum (black lines)
and the ion kinetic energy power spectrum (red lines) for models A1, A3 and A6.
Each model has its own panel from left to right, respectively.   Symbols indicate the full kinetic energy spectrum
while lines indicate the energy spectrum of the Alfv\'en modes only.
We modify the power spectrum by $k^{5/3}$, which is the predicted
slope for both Alfv\'enic turbulence and Kolmogorov-type turbulence.}
\label{fig:var_lin}
\end{figure*}

While this work is a theoretical and numerical study, our results have very important direct implications for observations.
We have shown that the Alfv\'en Mach number is a critical parameter for obtaining the damping and scaling characteristics of the
turbulence which may play a role in the interpretation of velocity dispersion methods of obtaining the ambipolar diffusion scale (Hezareh et al. 2014).
Our results fit well in the context of interpreting the observational signatures of decoupling found in Tilley \& Balsara (2010) and Meyer et al. (2014). Meyer et al. (2014) showed that the PDFs of sub-Alfv\'enic turbulence (i.e. models A3 and A6)  were different in the ions and neutrals when
looking at a LOS parallel to B, but similar with a sight-line perpendicular to B.  As we have shown, the
perpendicular decoupling scale for these models occurs at much smaller scales along the cascade, hence density fluctuations maybe coupled in this direction
even after the cascade has begun to damp.  Contrary to the sub-Alfv\'enic behavior, 
Meyer et al. 2014 showed that the PDFs of ions and neutrals
for model A1 are very similar and are generally LOS independent, which is expected
from our results as turbulence persists in this model to scales smaller than the decoupling scale.

One of our most important findings suggest that the MHD turbulence cascade does not simply damp at the decoupling scale L$_{AD}$ but rather depends on other parameters
such as the Alfv\'en Mach number, which determines the level of anisotropy in the cascade.  Furthermore, we show that neutrals and ions behave as a single MHD fluid at certain scales, which can explain the correlation of 21-cm filaments with the magnetic field (Clark, Peek, \& Putman 2014). There are several methods
to obtain the Alfv\'en Mach number that extend beyond direct observational techniques such as Zeeman spiting (Crutcher et al. 2009).
These include anisotropy in the structure function of velocity centroids (Esquivel \& Lazarian 2005; Esquivel \& Lazarian 2011; Burkhart et al. 2014), 
Principle Component Analysis (Correia et al. 2014) and the bispectrum (Burkhart et al. 2009).  The power spectrum of partially ionized fluids can also be an indication of the Alfv\'enic state of the gas, as suggested by Hezareh et al. (2010) and Meyer et al. (2014) and shown in our Figure \ref{fig:var_lin}.

Our work shows that the use of scale $L_{AD}$ as the
scale for the suppression of the MHD cascade in partially ionized gasses may not be the final story. It is clear that a more detailed theoretical study of turbulence dissipation in partially ionized gas is required in order to understand the precise damping mechanism for a given parameter space (see more in Xu et al. 2014, in prep.)

\section{Conclusions}
\label{sec:con}

We studied the structure function scaling relations 
and power spectra of the velocity field Alfv\'en modes of two fluid (ion-neutral) MHD turbulence.
We investigated at what scales MHD turbulence is damped in the partially ionized gas, treating ions and neutral separately.
We showed that the GS95 scalings and anisotropy are present in the velocity structure functions of the ions and neutrals in all models studied however the dynamic range of scales over which GS95-like turbulence is seen varies based on Alfv\'en Mach number and the
range of the parallel and perpendicular ambipolar diffusion scales bounded by the viscous dissipation scale and the driving scale. We found that 
Alfv\'enic turbulence does not necessarily damp at the ambipolar diffusion scale in super-Alfv\'enic turbulence and that the damping depends on other turbulence parameters such as the Alfv\'en Mach number and driving scale. We also showed that the power spectrum of the kinetic energy in the damped regime is consistent with a $k^{-4}$ scaling, as predicted by LVC04.

\acknowledgments
B.B. acknowledges support from the NASA Einstein Fellowship.
A.L. and B.B. thank the Center for Magnetic Self-Organization in Astrophysical and Laboratory Plasmas for financial support
and acknowledge financial support of the INCT INEspa�o and the Physics Graduate Program/UFRN, at Natal, for hospitality.
AL is
supported by the NSF grant AST 1212096.
DSB acknowledges support via NSF grants NSF-AST-1009091 , NSF-ACI-1307369 and NSF-DMS-1361197. DSB also acknowledges support via NASA grants from the Fermi program as well as NASA-NNX 12A088G. Several simulations were performed on a cluster at UND that is run by the Center for Research Computing.  Computer support on NSF's XSEDE computing resources is also acknowledged.

\end{document}